\documentclass[12pt]{iopart}
\usepackage{graphicx}
\usepackage{epsfig}
\begin{document}
\newcommand{\norm}[1]{\ensuremath{| #1 |}}
\newcommand{\aver}[1]{\ensuremath{\langle #1 \rangle}}
\newcommand{\ket}[1]{\ensuremath{| #1 \rangle}}
\newcommand{\bra}[1]{\ensuremath{\langle #1 |}}
\newcommand{\braket}[2]{\ensuremath{\langle #1 | #2 \rangle}}
\newcommand{\imag}{{\rm i}}

\title{Excitations in two-component Bose-gases}

\author{A Kleine$^1$, C Kollath$^{2,3}$, I P McCulloch$^1$, T Giamarchi$^2$ and U Schollw\"ock$^1$}

\address{$^1$ Institute for Theoretical Physics C, RWTH Aachen,
 D-52056 Aachen, Germany}

\address{$^2$ DPMC-MaNEP, University of Geneva, 24 Quai
 Ernest-Ansermet, CH-1211 Geneva, Switzerland }

\address{$^3$ Centre de Physique Th\'eorique, Ecole Polytechnique, 91128
 Palaiseau Cedex, France }

\eads{\mailto{kleine@physik.rwth-aachen.de}, \mailto{kollath@cpht.polytechnique.fr}}

\begin{abstract}
In this paper, we study a strongly correlated quantum system that has become amenable to experiment by the advent of ultracold bosonic atoms in optical lattices, a chain of two different bosonic constituents. Excitations in this system are first considered within the framework of bosonization and Luttinger liquid theory which are applicable if the Luttinger liquid parameters are determined numerically. The occurrence of a bosonic counterpart of fermionic spin-charge separation is signalled by a characteristic two-peak structure in the spectral functions found by dynamical DMRG in good agreement with analytical predictions. Experimentally, single-particle excitations as probed by spectral functions are currently not accessible in cold atoms. We therefore consider the modifications needed for current experiments, namely the investigation of the real-time evolution of density perturbations instead of single particle excitations, a slight inequivalence between the two intraspecies interactions in actual experiments, and the presence of a confining trap potential. Using time-dependent DMRG we show that only quantitative modifications occur. With an eye to the simulation of strongly correlated quantum systems far from equilibrium we detect a strong dependence of the time-evolution of entanglement entropy on the initial perturbation, signalling limitations to current reasonings on entanglement growth in many-body systems.
\end{abstract}
\maketitle
%%%%%%%%%%%%%%%%%%%%%%%%%%%%%%%%%%%%%%%%%%%%

\section{Introduction}
One of the key proposals in the field of quantum computing, made by
Feynman, is the idea to use one quantum system to simulate another one
in order to circumvent the problem of the qualitatively different
complexity of quantum systems and classical computers as the standard
simulation tools of science. The advantage of Feynman's approach lies 
in the fact that the new simulating quantum system may have advantages
in experimental control: preparation, manipulation of parameters and
measurement.

Over the last decade, this proposal has been filled with life by the
progress made in the
preparation of dilute ultracold atom gases, highlighted by the now
almost routinely achieved preparation of Bose-Einstein condensates.
The use of Feshbach resonances or optical lattices has given us 
unprecedented control over interaction and dimensionality in
strongly interacting quantum many-body systems of unique purity. This 
has been put to use in the creation of strongly correlated quantum
systems that have been in the focus of interest in condensed-matter
physics for a long time\cite{BlochDalibardZwerger2007}. Examples are the
observation of the superfluid to Mott insulator transition for Bose gases
\cite{GreinerBloch2002} and the fermionization of strongly interacting
one dimensional bosons\cite{ParedesBloch2004,KinoshitaWeiss2004}.

But one can do more: it is possible to create physical systems of
their own interest that have no counterpart in conventional condensed 
matter physics. To take one example, due to the internal spin degree 
of freedom of electrons it is quite natural in solids to consider 
models of two components of fermions. The existence of two fermionic 
components is the driving force of important phenomena such as 
collective magnetism.

In this paper, we consider the bosonic equivalent of the two-component 
fermionic system. Strongly interacting two-component bosonic systems 
have no counterpart in condensed-matter physics, which is why they 
have found limited attention in the solid-state literature.
Yet, they are quite 
easily implemented in the field of ultracold atom gases\cite{ErhardSengstock2004,WideraBloch2004}, and have generated quite some interest (for a review, see \cite{Lewenstein2007}). We focus on 
the case of one-dimensional two-component bosonic systems: on the one 
hand, they offer interesting physical phenomena such as a bosonic
version\cite{CazalillaHo2003,ParedesCirac2003,KleineSchollwoeck2007} of fermionic spin-charge separation which was discussed for cold atoms by \cite{RecatiZoller2003,KeckeHaeusler2004,KollathZwerger2005,KollathSchollwoeck2006b}. On the other hand, very good 
control exists both analytically via bosonization\cite{Giamarchibook} and numerically via 
(time-dependent) DMRG\cite{White1992,Schollwoeck2005}. 

In this paper, we start out by considering the low-energy physics
which is characterized, as all other critical one-dimensional quantum 
systems, by a few effective Luttinger liquid parameters, which we 
determine both by a mapping in a limiting regime and more generally by DMRG. 
In principle, the Luttinger liquid parameters 
determine completely the static and linear response properties of the 
system.

We move on to discuss the spectral functions which are 
the cleanest way to observe Luttinger liquid physics, in particular spin-charge separation and characterize the linear-response behaviour of two-component bosonic systems. These are obtained in the framework of dynamical DMRG.

Experimentally, one will be confronted by certain limitations of ultracold atomic systems: currently, spectral functions are unavailable and one is restricted to monitor the time-evolution of excitations. These in fact show a separation of symmetric and antisymmetric density combinations (``charge" and ``spin") of the individual species from which results in good agreement with those from the spectral functions can be derived. These observation still hold if we also take into account that in current implementations of two-component bosonic systems there are slight differences in the intraspecies interactions of the two components. Moreover, the presence of a confining harmonic trap potential does not qualitatively alter the results, as we can show by explicit simulation using time-dependent DMRG.

Last, but not least, we turn to the discussion of the time-evolution of the entropy of entanglement in the various out-of-equilibrium scenarios considered in this paper. This study was originally motivated by the fact that the efficiency of DMRG simulations is limited by entanglement growth which impacts exponentially on the numerical resources needed. Here, it turns out that there are surprisingly large variations in the generally accepted scenario of an essentially linear entanglement growth after quenches (as which all our scenarios can be interpreted). While we cannot give a general explanation for the phenomenon, we hope to provide a stimulus for further research to develop a more complete understanding of entanglement evolution.

\section{Model}
Cold atomic gases with two hyperfine species confined in optical lattices can
be described by the two-component Bose-Hubbard model given by the following
Hamiltonian\cite{JakschZoller1998}:
\begin{equation}
\label{eq:H}
\begin{array}{rl}
H =& -J \sum_{j,\nu} \left( b^\dagger_{j+1,\nu}b_{j,\nu} +
h.c.\right) + \sum_{j,\nu} \frac{U_\nu \hat{n}_{j,\nu}(\hat{n}_{j,\nu}-1)}{2}  \\
&+U_{12} \sum_j \hat{n}_{j,1}\hat{n}_{j,2}+ \sum_{j,\nu}
\varepsilon_{j,\nu} \hat{n}_{j,\nu} ,
\end{array}
\end{equation}
where $j$ is a site index and $\nu$ labels the two different flavours of bosons. $J$ is the hopping strength, $U_\nu$ the intraspecies and $U_{12}$
the interspecies onsite interaction. Here we assume that the two hyperfine
species have the same mass and see the same lattice potential. Unless otherwise stated we use $U_1 = U_2 \equiv U$, and $u = U / J$, $u_{12} = U_{12} / J$.
$\varepsilon_{j,\nu}$ describes an external potential, i.e. given by a trap. \\
In the following we denote the lattice spacing by $a$.
We mainly consider incommensurable fillings with equal densities $n_1 = n_2 \equiv n$ smaller than one. For vanishing interspecies interaction we recover the
one component Bose-Hubbard model being in a superfluid phase. The superfluid phase remains stable for finite $U_{12}$ up to $U_{12} \sim U$. For $U_{12} > U$
the interspecies interaction becomes dominant and a demixing of the flavours occurs, i.e. a phase separation\cite{CazalillaHo2003,MishraDas2006}. The physics corresponds then to the one of an itinerant ferromagnetic system with non-Luttinger liquid properties \cite{Zvonarev2007,FuchsShlyapnikov2005}. Experimentally, to test for Luttinger liquid properties, the relevant regime is given by $U_{12} \sim U$, with $U_{12}$ slightly below $U$ to avoid the phase-separation regime. Strongly differing interaction parameters have not been realized experimentally so far; our typical interaction parameters are chosen accordingly. 

\section{Approximations}
\paragraph{Continuum model}
In a weakly interacting superfluid phase in the low filling limit, the Bose-Hubbard model can be mapped
to the continuous Lieb-Liniger model \cite{LiebLiniger1963I,Lieb1963}. The mapping can be performed taking the
limit $a\to 0$ while leaving $Ja^2$ constant.  

 The Hamiltonian for two bosonic species in the continuum is
\begin{eqnarray}
\label{eq:LL}
&&H_{LL}=\int {\textrm d}{x} \;\sum_{\nu=1,2} \left(\frac{1}{2M}\norm{\partial_x \Psi_\nu(x)}^2  \right.\nonumber\\
&&+\left. V(x) \Psi_\nu^\dagger (x)\Psi_\nu(x)+ \frac{g}{2} (\Psi_\nu^\dagger (x))^2(\Psi_\nu(x))^2\right)\nonumber\\
&&+\frac{g_{12}}{2}\int {\textrm d}{x} \; (\Psi_1^\dagger (x)\Psi_1(x) )(\Psi_2^\dagger (x)\Psi_2(x)).
\end{eqnarray}
Here $\Psi^{(\dagger)}$ is the bosonic annihiliation (creation) operator, $V$
the external potential, $M$ the mass of the particles, and
$g$ and $g_{12}$ are the strengths of the intra- and inter- species interaction,
respectively. The parameters of the continuum model and the lattice model are
related by 
$Ja^2=\frac{1}{2M}$, the interaction strength to the
$\delta$-interaction strength by $Ua=g$ and $U_{12}a=g_{12}$, and the density $\rho$ to the
filling factor $n$ by $\rho a = n$.
In the hydrodynamic
approximation \cite{PitaevskiiStringari2003} the sound velocities of this model are given
by
\begin{eqnarray}
v_{c,s}&&=v_{0}/\sqrt{1\pm g_{12}/g}\\
\textrm{with}\quad v_0&&= \sqrt{g \rho /M}.
\end{eqnarray}
The indices stand for charge and spin, corresponding to symmetric (charge) and antisymmetric (spin) combinations of the two bosonic fields.
At the symmetric point, where the interspecies interaction is equal to the
intraspecies interaction, the model is SU(2) symmetric and can be solved by the
Bethe ansatz. For the special case $g_{12}=g$ the sound velocity has been determined using
the Bethe ansatz \cite{FuchsShlyapnikov2005, BatchelorOelkers2006}. The spin
dispersion becomes quadratic at this point. 

\paragraph{Bosonization}
The low energy physics of the Bose-Hubbard model can be described by the
bosonization approach\cite{Giamarchibook}. Two bosonic fields, $\theta_\nu$ and $\phi_\nu$, are
introduced in the continuum which are related to the phase and the amplitude of the original
bosonic operator, respectively. To be more precise, in this representation the bosonic creation
operator becomes $b_\nu^\dagger(x)= \sqrt{\rho_0} \sum_p \e^{\imag 2 p (\pi \rho_0
 x-\phi_\nu(x))} \e^{-\imag \theta_\nu}$ and the density operator $ \rho_\nu(x) =\rho_0
-1/\pi \nabla \phi_\nu(x) + \rho_0 \sum_{p\not = 0} \e^{\imag 2 p (\pi \rho_0
 x-\phi_\nu(x))}$. Here $\rho_0$ is the average
density, and $\frac{1}{\pi} \partial_x\phi_\nu$ and $\theta_\nu$ are conjugate
operators.
The advantage of this representation is that the low-energy properties of the system are described by a quadratic Hamiltonian. Introducing the ``charge" and ``spin" degrees of freedom by 
$\phi_c=1/\sqrt{2} (\phi_1+\phi_2)$ and $\phi_s=1/\sqrt{2} (\phi_1-\phi_2)$ it
separates into two different part and is given by
\begin{eqnarray}
\label{eq:Hbos}
H&=&H_c +H_{s} \quad \textrm{with}\\
H_c& =& \frac{1}{2\pi} \int \textrm{d}x \; \left[ v_c K_c (\partial \theta_c)^2 + \frac{v_c}{K_c}
(\partial_x \phi_c)^2 \right] \; \textrm{and} \nonumber\\\\
H_s& =& \frac{1}{2\pi} \int \textrm{d}x\; \left[ v_s K_s (\partial
\theta_s)^2 + \frac{v_s}{K_s}(\partial_x \phi_s)^2 \right] \nonumber\\&+& \frac{g_{12}}{(\pi
\rho^{-1})^2 }\int \textrm{d}x\; \cos (\sqrt{8\phi_s}).
\end{eqnarray}
Here the parameters $v_\nu$ are the velocities and $K_\nu$ the Luttinger liquid
parameters for the corresponding field. Due to this mapping the asymptotic
physics of the Bose-Hubbard model is totally determined by the velocities and
the Luttinger liquid parameters. 
In the case of intermediate interaction strengths, the relations between the microscopic parameters of the Bose-Hubbard model and the parameters of the bosonization approach are difficult to establish analytically. In the limit of small interaction strength and low filling they are given for the single
species Bose-Hubbard by $Jan \pi=v_0 K$ and $Ua/2\pi=v_0/K$. 

In the limit of weak interspecies coupling the parameters for charge and spin
degrees of freedom can be related to the parameters of the system without interspecies
coupling by the perturbative expressions
\begin{equation}
\label{eq:bosonization}
\begin{array}{rl}
v_{c,s} = & v_0 \sqrt{1 \pm \frac{g_{12} v_0}{\pi K_0}} \\
K_{c,s} = & K_0 / \sqrt{1 \pm \frac{g_{12} v_0}{\pi K_0}}.
\end{array}
\end{equation} 
Outside the range of validity of these two approximations the parameters can be
determined numerically which we do in the following for a broad range of parameters. The Luttinger parameter is determined by using
\begin{equation}
\label{eq:K_kappa_v}
K_c =\frac{\pi}{2} ~ v_c ~ \kappa_c~, \quad K_s = \frac{\pi}{2} ~ v_s ~ \kappa_s.
\end{equation} 
The compressibility \cite{Giamarchibook} is given by $\kappa_{c,s} = \left(
\frac{\partial^2}{\partial n_{c,s}^2} \frac{E_0}{L} \right)^{-1}$, where
$E_0(L)$ is the energy of the ground state for a system of length $L$
and the combined  ``charge" and ``spin" densities are denoted $n_{c,s}=n_1 \pm n_2$. The derivative can be computed numerically, yielding
\begin{equation}
\label{eq:kappa}
\begin{array}{rl}
\kappa_{c}^{-1} \approx & L \frac{ E(N+\Delta N,N+\Delta N) + E(N-\Delta N,N-\Delta N) - 2 E(N,N) }{4 \Delta N^2} \\
\kappa_{s}^{-1} \approx & L \frac{ E(N+\Delta N,N-\Delta N) + E(N-\Delta N,N+\Delta N) - 2 E(N,N) }{4 \Delta N^2}.
\end{array}
\end{equation} 
$E(N_1,N_2)$ is the ground state energy of the system with $N_1$ particle of the first flavour and $N_2$ of the second one.
After a $L \to \infty$ extrapolation, we can obtain the Luttinger parameter by using Eq.~(\ref{eq:K_kappa_v}) and the numerically determined velocities. Note that the formulae given for $K_s$ and $\kappa_s$ in Equations (\ref{eq:K_kappa_v}) and (\ref{eq:kappa}) differ by (mutually cancelling) factors of 4 from the usual relationships in electronic models, where $n_s=(n_1 - n_2)/2$; the factor $1/2$ is not meaningful in the model considered here and hence dropped. The final value of $K_s$ is of course unaffected.\\
 
\begin{figure} 
\begin{center}
{\epsfig{figure=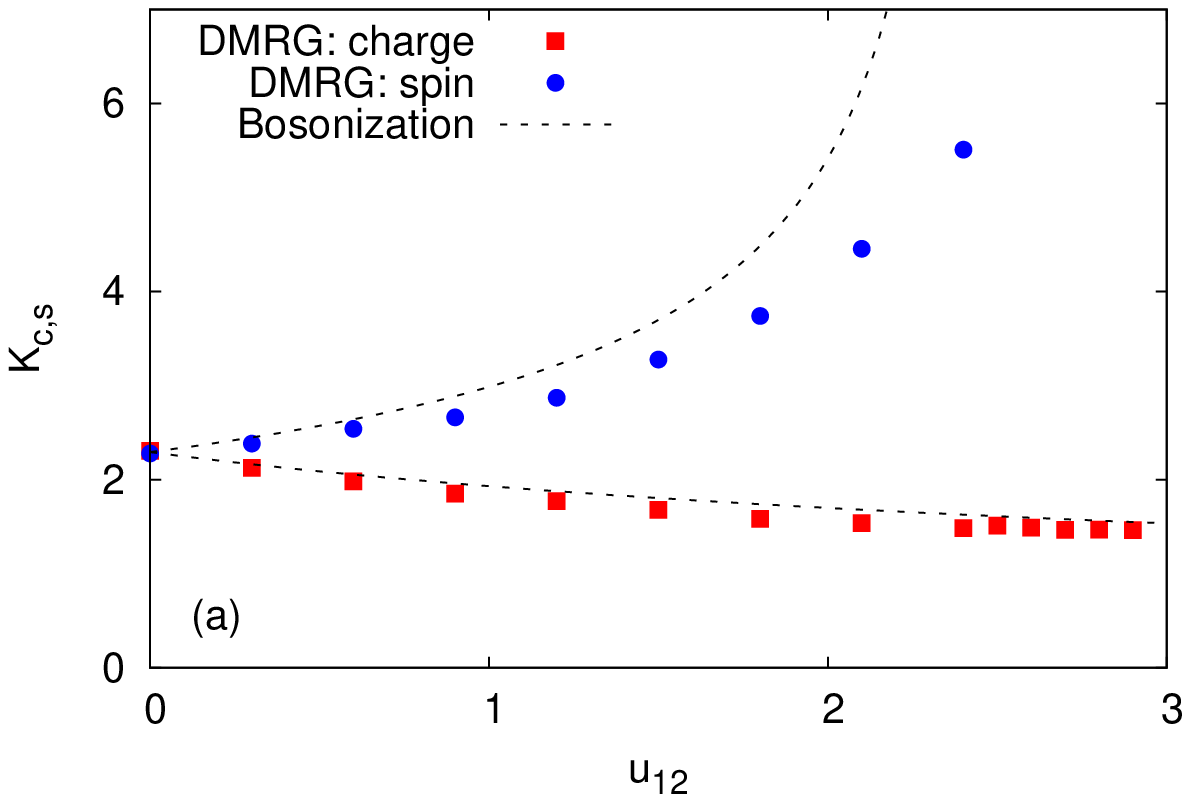,width=0.5\linewidth}}\\
{\epsfig{figure=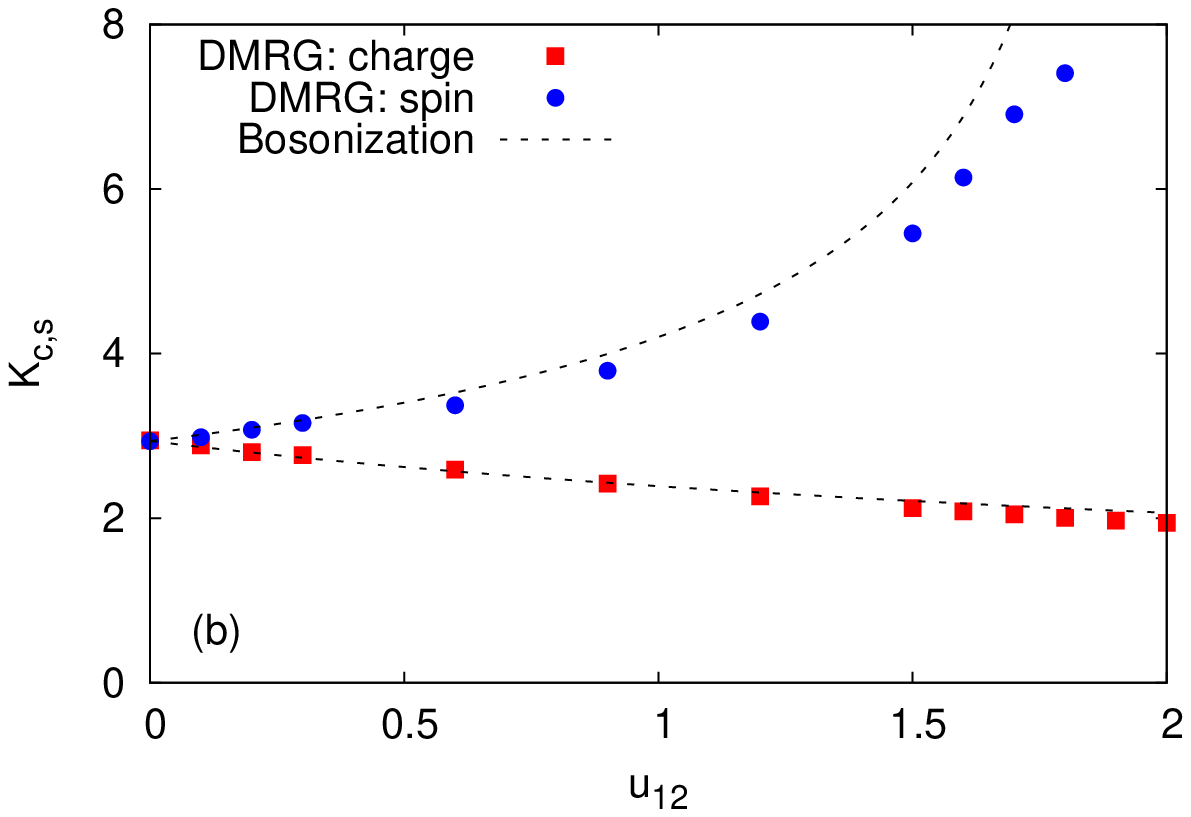,width=0.5\linewidth}}
\end{center}
\caption{(Color online) Dependence of the charge and spin Luttinger parameter on the
 interparticle interaction strength $u_{12}$. A comparison of analytical results (line, see text) and
 numerical DMRG results (symbol) is shown. The parameters used are (a) $u=3$, $n\approx 0.63$ and (b) $u=2$, $n\approx 0.88$.}
\label{fig:Lutt}
\end{figure}

\begin{figure} 
\begin{center}
{\epsfig{figure=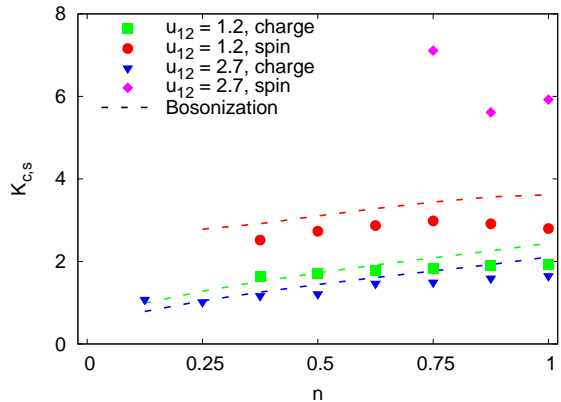,width=0.5\linewidth}}
\end{center}
\caption{(Color online) Dependence of the charge and spin Luttinger parameter on the
charge background density. A comparison of analytical results (line, see text) and
numerical DMRG results (symbol) is shown. The parameters used are (a) $u=3$,
$u_{12}=1.2$ and (b) $u=3$, $u_{12}=2.7$.}
\label{fig:Lutt_n}
\end{figure}

Fig.~\ref{fig:Lutt} shows the dependence of the Luttinger parameters on the interspecies interacting strength $u_{12}$.
The charge Luttinger parameter $K_c$ decreases with increasing $u_{12}$, the spin Luttinger parameter increases.
As expected, the bosonization result agrees with the
numerical results for small $u_{12}$ and starts to deviate with increasing $u_{12}$. 
The spin Luttinger parameter shows stronger deviations. Fig.~\ref{fig:Lutt_n} shows the Luttinger parameter for fixed interaction strength ($u=3, u_{12} = 1.2,~ 2.7$) and
varying density $n$. Note that for the spin Luttinger parameter $u_{12} v_0 / (\pi K_0)$ is of the order of one, and thus the perturbative result (Eq.~\ref{eq:bosonization})
is not reliable anymore. The DMRG results deviates significantly from the perturbative results; the deviations increase with density. A similar effect also occurs in the
one-component Bose-Hubbard model where for densities above $0.5$ lattice effects becomes relevant, see the Appendix for more details. \\

\section{Spectral Functions}
The key quantity to characterize single particle
excitations in many-body systems is the dynamic single-particle spectral function, because it can be probed easily in solid-state setups. The single particle spectral function is defined as
\begin{equation}
\label{eq:spectral_func}
A(q,\omega+i\eta) = - \frac1{\pi} \Im \bra{0} b_{q,1} \frac1{E_0 + \omega +i \eta - \hat{H}} b^\dagger_{q,1} \ket{0},
\end{equation}
where $\ket{0}$ is the ground state with energy $E_0$. 
For fermionic two-component Hubbard model this function signals the separation
of spin and charge degrees of freedom by two distinct peaks at different frequencies.

In the following we determine this function for the two-species bosonic model using a
variant of dynamical DMRG\cite{KuehnerWhite1999,Jeckelmann2002}. Our formulation of the algorithm is entirely based on matrix product states. This allows to avoid the targetting of multiple states in the DMRG algorithm, which is achieved only at substantial numerical cost. The new formulation of the algorithm saves more than an order of magnitude in time. As always, DMRG prefers open boundary conditions leading to spurious effects in the spectral functions; in this context, filtering procedures have been used\cite{KuehnerWhite1999}. In order to reduce boundary effects we use the quasi-momentum definition of the Fourier transform
$ b_{q,\nu} = \sum_j \sin\left( \frac{qj\pi}{L+1} \right) b_{j,\nu} $ with the momentum $|k| = q \pi / (L+1)$.
These states create a single particle basis of eigenstates of the non-interacting system ($u = u_{12} = 0$) with open boundary conditions. They have nodes at the edges of
the systems and are therefore better suited to open boundary conditions than the usual definition of the Fourier transform using the eigenfunctions of the non-interacting
system with periodic boundary conditions. \\

Dynamical DMRG uses a correction vector method to calculate the spectral function: First define $ \ket{lv(q,\omega)} = b^\dagger_{q,1} \ket{0}$. Then the correction vector
is given by $ \ket{cv(q,\omega+i\eta)} = \frac1{E_0 + \omega +i \eta - \hat{H}} \ket{lv(q,\omega)} $ and thus obey 

\begin{equation}
\label{eq:cv}
\left( E_0 + \omega +i \eta - \hat{H}\right) \ket{cv(q,\omega+i\eta)} = \ket{lv(q,\omega)}.
\end{equation}

This complex equation for the correction vector can then be solved using the GMRES algorithm. The value of the spectral
function is thus $ A(q,\omega) = - \frac1\pi \Im \braket{lv(q,\omega)}{cv(q,\omega+i\eta)} $. A significant improvement is gained by considering the derivative
of $A(q,\omega+i\eta)$,
\begin{eqnarray}
\frac\partial{\partial \omega} A(q,\omega+i\eta) & = &
\frac1\pi \Im \bra{lv(q,\omega)} \left(E_0 + \omega +i \eta - \hat{H}\right)^{-2}  \ket{lv(q,\omega)} \nonumber \\
& = & \frac1\pi \Im \braket{cv(q,\omega-i\eta)}{cv(q,\omega+i\eta)} \nonumber \\
& = & \frac1\pi \Im \braket{\overline{cv}(q,\omega+i\eta)}{cv(q,\omega+i\eta)}
\end{eqnarray}
where the last line is only valid if $ \ket{lv(q,\omega)} = \ket{\overline{lv}(q,\omega)} $ and $\ket{\overline{\Psi}}$ denotes the complex conjugate of $\ket{\Psi}$.
Using the derivative, we can determine the spectral function with a smaller number of correction vectors and thus much more efficiently.\\

\begin{figure} 
\begin{center}
       {\epsfig{figure=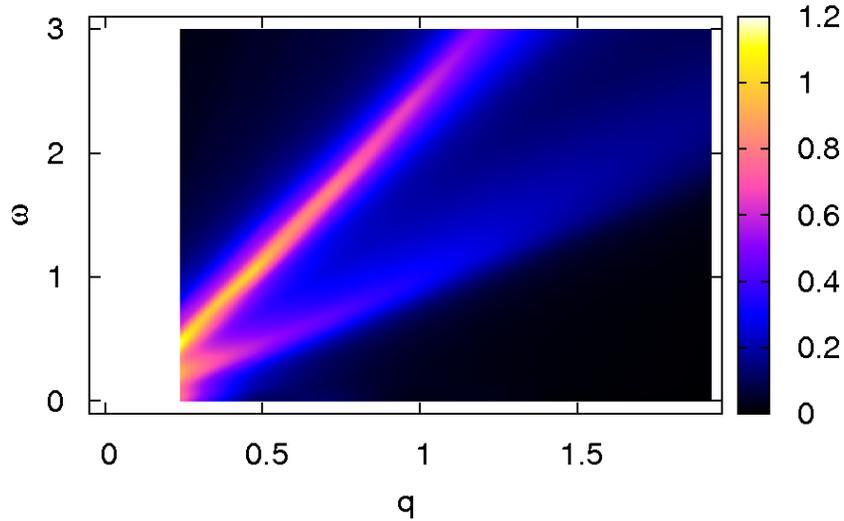,width=0.8\linewidth}}
\end{center}
\caption{(Color Online) Density plot of the one-particle spectral functions $A(q,\omega)$. The following parameters
 were used $n = 0.625$, $u = 3$, $u_{12} = 2.1$ on a system with
 $L=64$ sites and a broadening $\eta=0.1$. The charge branch is above the spin branch.}
\label{fig:spectral_density}
\end{figure}

\begin{figure} 
\begin{center}
       {\epsfig{figure=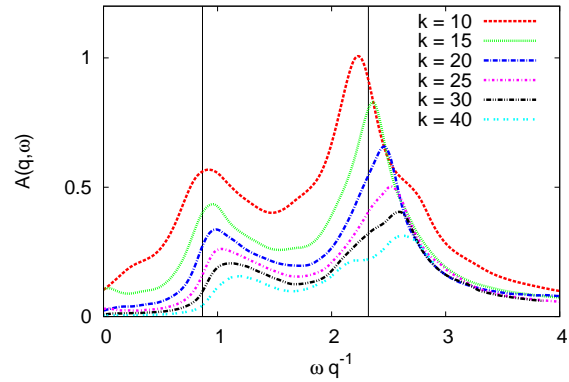,width=0.5\linewidth}}
\end{center}
\caption{(Color online) One-particle spectral functions $A(q,\omega)$ at momenta $q = k/65 \pi/a$ plotted against $\omega q^{-1}$.
 Two peaks corresponding to the spin
 and the charge excitation can be distinguished. The vertical lines mark the
 position of $u_{c,s}$. The following parameters
 were used $n = 0.625$, $u = 3$, $u_{12} = 2.1$ on a system with
 $L=64$ sites and a broadening $\eta=0.1$. }
\label{fig:spectral}
\end{figure}

Note that in the following only the normalized spectral function
\begin{equation}
A_{norm}(q,\omega) = A(q,\omega) \Big( \braket{lv(q,\omega)}{lv(q,\omega)} \Big)^{-1}
\end{equation}
is used, such that $\int d\omega ~ A_{norm}(q,\omega+i\eta) = 1 $ holds for every $q$.
The full spectral function $A_{norm}(q,\omega)$ is shown in Fig.~\ref{fig:spectral_density}. For the used system parameters $L = 64$ and $\eta = 0.1$ we needed up to
2000 states for each correction vector. One clearly sees two different branches with a linear dispersion relation $\omega \approx v_{c,s} q$ yielding two different velocities.
Thus this system exhibits spin-charge separation. Fig.~\ref{fig:spectral} gives a more detailed view of the
spectral function. For a number of given momenta $q = k \pi / (L+1)$ the spectral function is plotted versus $\omega q^{-1}$. Therefore the norm of the scaled spectral function
obeys $\int d\left(\omega q^{-1}\right) ~ A_{norm}(q,\omega+i\eta) = q^{-1} $ and the norm decreases with increasing $q$. The position of the peak is roughly given by
$\omega \approx v_{c,s} q$, thus one expects two peaks at the two velocities $v_c$ and $v_s$\cite{IucciGiamarchi2007}. The spectral function shows a number of boundary effects similar to the one encountered
for a single-component Bose-Hubbard model, see the Appendix for a detailed discussion of the one-component model. Here an addition peak at $\omega \approx 0$ occurs, visible for instance for
$q = 15\pi/65$. Furthermore,  for small values of $q$ the charge peak is shifted to lower frequencies and above the charge peak an additional shoulder appears (i.e. for $q = 10\pi/65$). There
are also some effects which do not occur for a single-component model: for large momenta the charge peak splits up and an additional peak below the main peak emerges. The
plot also shows some effect of a non-linear dispersion. Both spin and charge peak are shifted to higher frequencies for higher momenta, similar to the effects visible for a Bose-Hubbard model,
see Fig.~\ref{fig:spectral_BH}. In Fig.~\ref{fig:spectral_density} momenta lower than $q \approx 0.25$ are not plotted, because for the system sizes considered, finite size effects dominate there and obscure the physically relevant information.

\section{Density perturbations and single-particle excitations}
In ultracold atom experiments, spectral functions are hard to observe, but using suitably tuned and focused lasers it is easy to create local density perturbations. Theoretically the
density perturbations can be created applying an external potential of the
form $\epsilon_{j,cs} \sim \exp ( - \left( j - j_0\right)^2/2\sigma_j ) $ for times $t < 0$. For times $t > 0$ the potential is switched off.

Time-evolutions can be calculated by adaptive time-dependent DMRG most easily by using Trotter decompositions of short-ranged Hamiltonians leading to local time-evolutions\cite{Vidal2004,DaleyVidal2004,WhiteFeiguin2004}. For longer-ranged interactions or systems with large local state spaces, it is more efficient to consider global time-evolutions\cite{Feiguin2005}. As we are dealing locally with products of two bosonic state spaces, we calculated the time-evolution of the density perturbations numerically in the latter framework. The algorithm was formulated using matrix product states and the global time evolution was done using a Krylov algorithm for exponentiation\cite{HochbruckLubich1997} with a fixed error bound per timestep \cite{McCulloch2008}.
The used error bound $\left\| \ket{\Psi(t+\Delta t)} - \exp(-i\hat H \Delta t) \ket{\Psi(t)} \right\|^2$ is of the
order of $10^{-5}$ with a timestep of $\Delta t \approx 0.1 ~ - ~ 0.2$. Usually, $6$ to $10$ Krylov vector were used.
For the Krylov vectors up to 3000 states were used in the case of the density perturbations.
The time-evolutions of single particle excitations are much harder to perform, up to 7500 states were used there.\\

\begin{figure} 
\begin{center}
   {\epsfig{figure=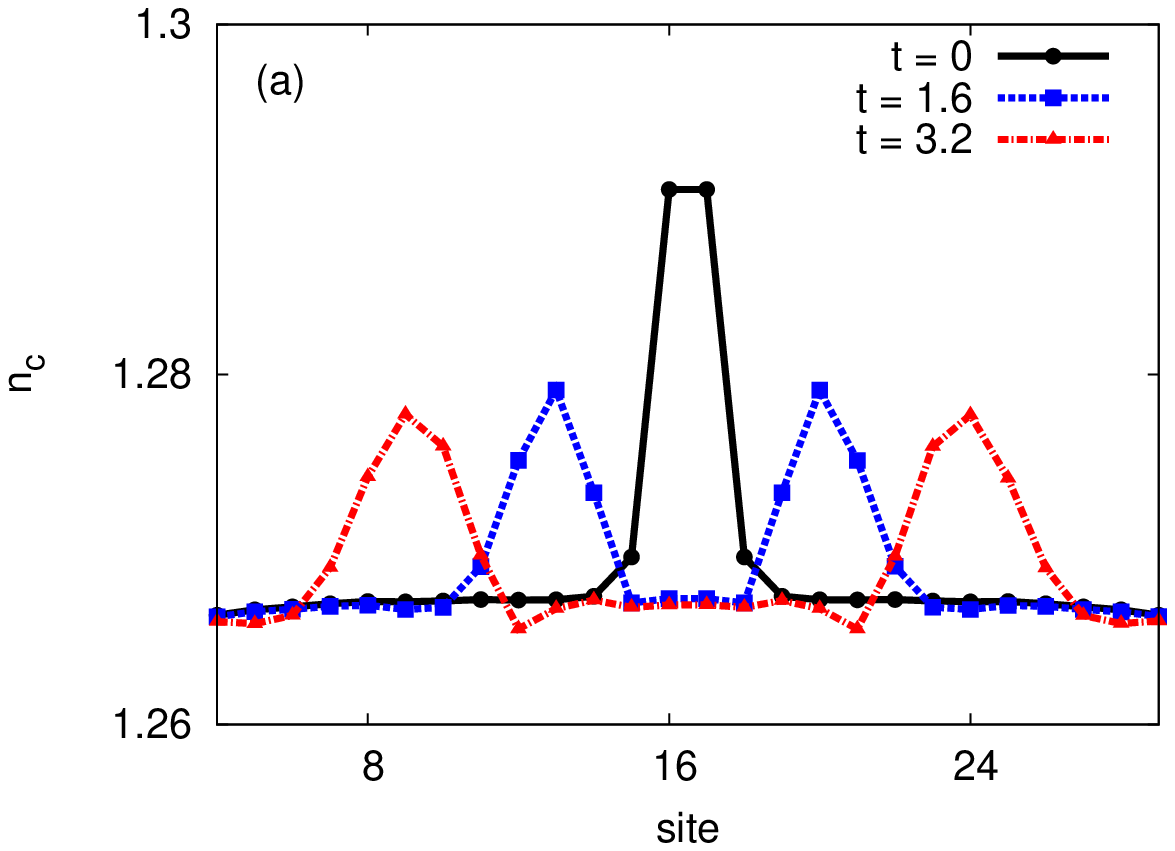,width=0.4\linewidth}}
   {\epsfig{figure=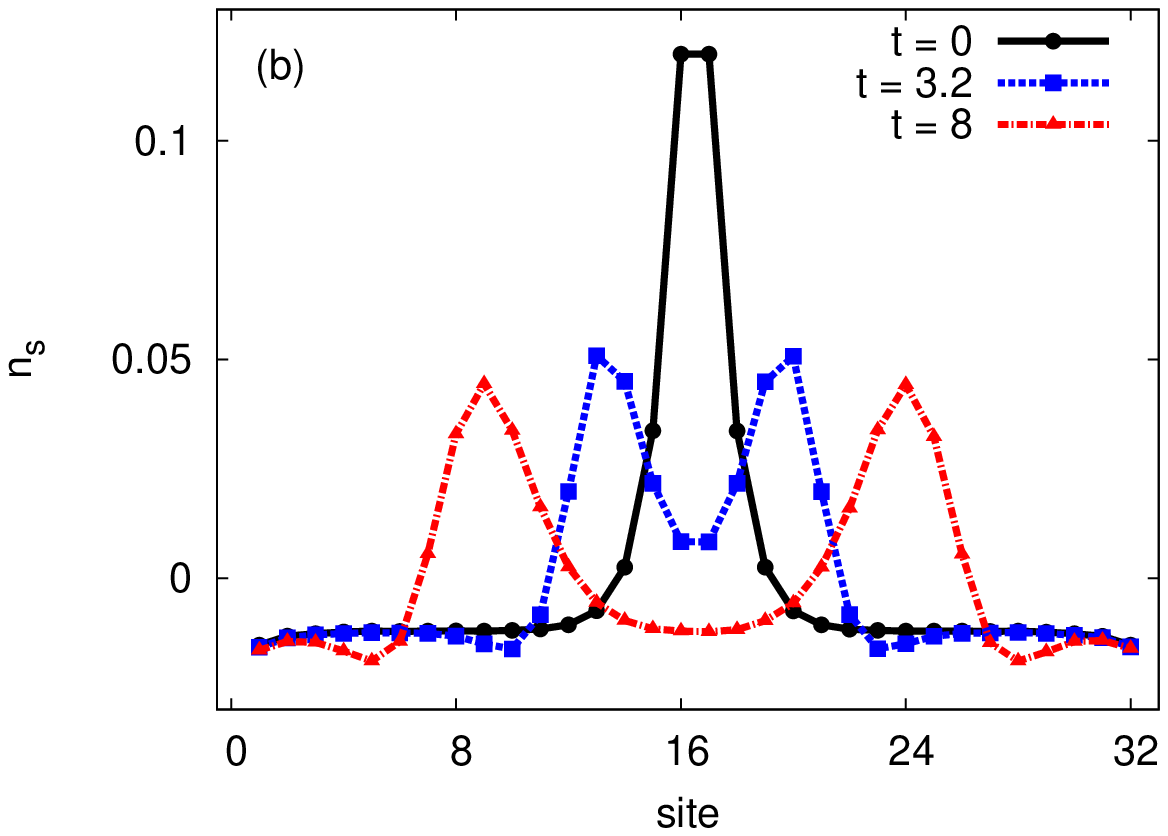,width=0.4\linewidth}}
\end{center}
\caption{(Color online) Snapshots of the time-evolution of the charge and spin density
 distribution of a small charge (a) and spin density perturbation (b) created at time $t=0\hbar/J$.
The system parameters are $n_{1,2} = 0.625$, $u = 3$, $u_{12} = 2.1$.}
\label{fig:smallperturbation_u2.1}
\end{figure}

\begin{figure} 
\begin{center}
   {\epsfig{figure=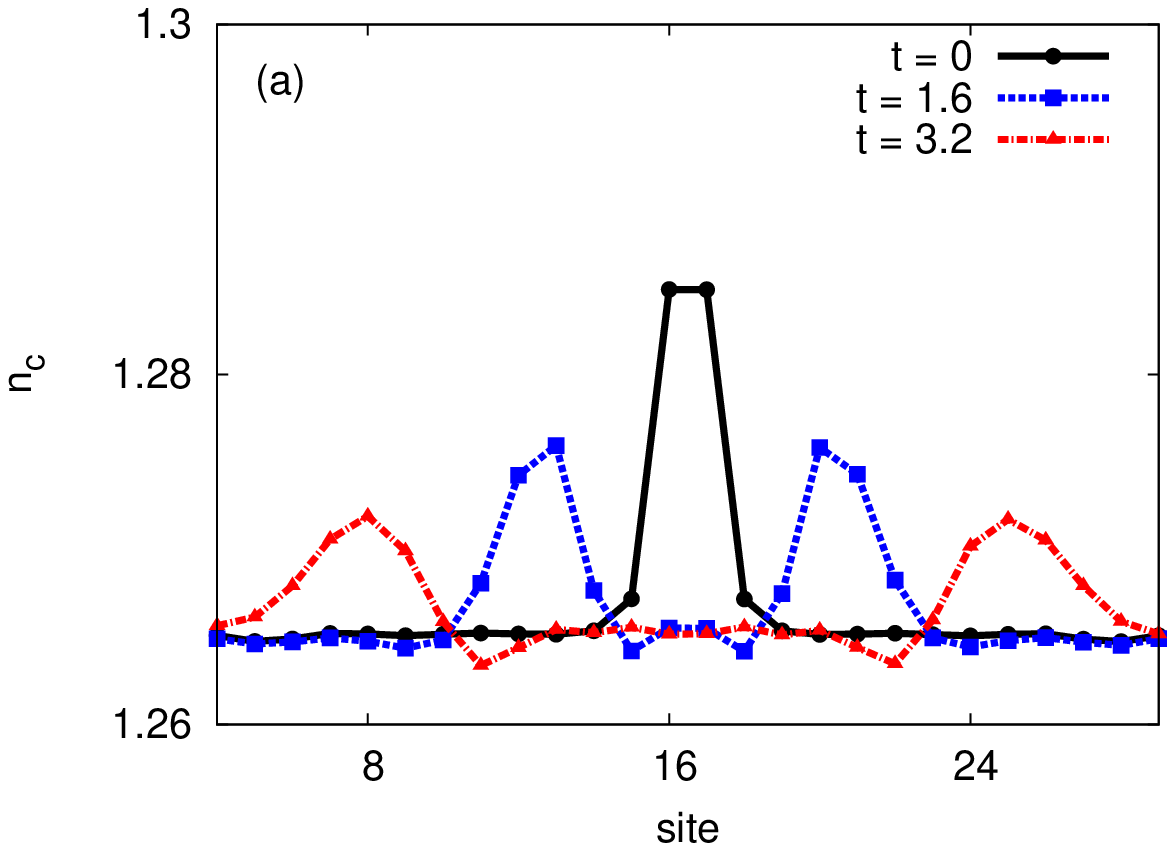,width=0.4\linewidth}}
   {\epsfig{figure=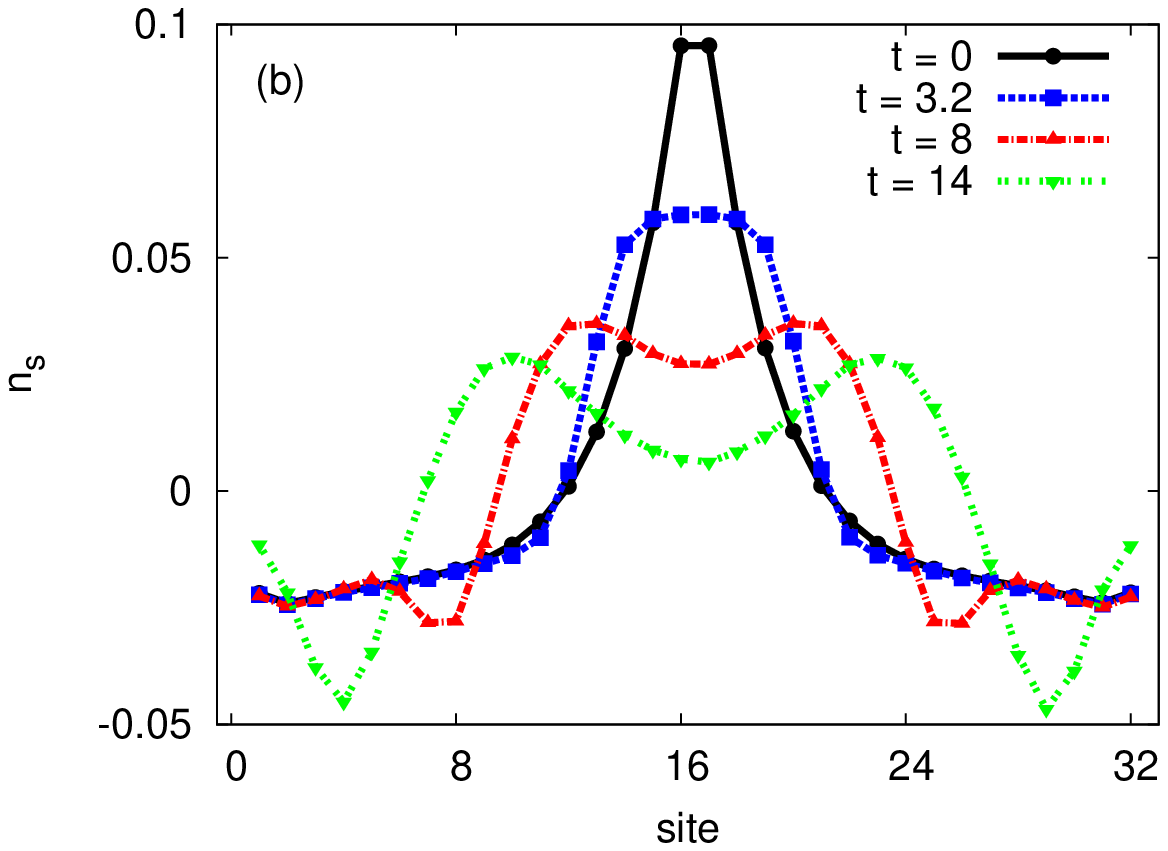,width=0.4\linewidth}}
\end{center}
\caption{(Color online) Snapshots of the time-evolution of the charge and spin density
 distribution of a small charge (a) and spin density perturbation (b) created at time $t=0\hbar/J$.
The system parameters are $n_{1,2} = 0.625$, $u = 3$, $u_{12} = 2.9$.}
\label{fig:smallperturbation_u2.9}
\end{figure}

Snapshots of the density evolution of these excitations for one parameter set are
shown in Fig.~\ref{fig:smallperturbation_u2.1}. One sees that the charge and
spin perturbation
which is created at time $t=0$ splits up into two counter-propagating
excitations. The velocity of the spin perturbation is much lower than the
velocity of the charge perturbation. Even after separating into two
perturbations the amplitude shows a decay. This decay is very slow for weak
inter-species interaction
(see~Fig.~\ref{fig:smallperturbation_u2.1}). However, if the inter-species
interaction is approaching the value of the intra-species interaction a strong
broadening of the spin perturbation can be seen (see~Fig.~\ref{fig:smallperturbation_u2.9}). Due to the
very low velocity and the broadening the two peaks only separate at very
long times. The strong broadening of the spin peak is
expected since in the limit of equal interaction strength the system has a
quadratic spin dispersion relation. Therefore the region of the linear regime of the
dispersion for $u_{12} <u$ decreases if $u_{12}$ gets closer to $u$. 
The dips in front of the spin perturbation and behind the charge perturbations
are due to the remaining interaction between the spin and the charge degrees
of freedom and finite-size effects. \\

\begin{figure} 
\begin{center}
   {\epsfig{figure=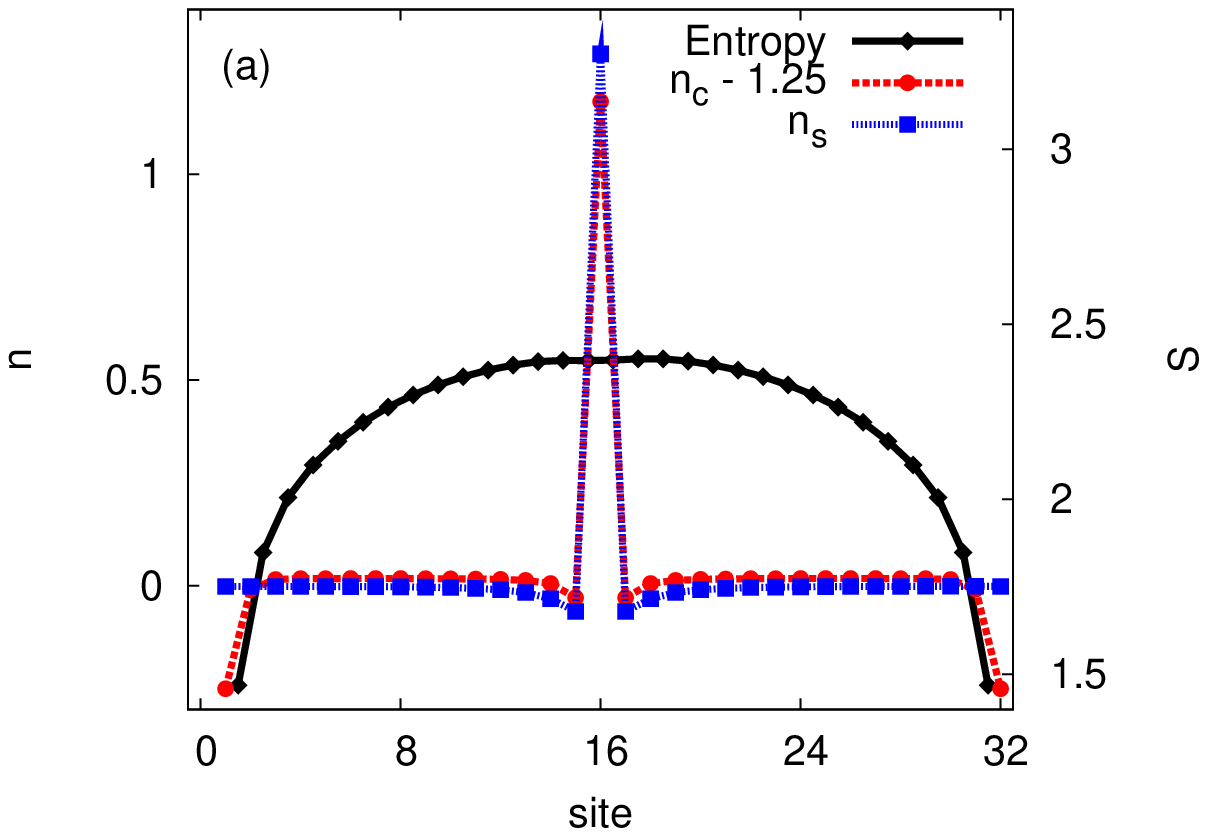,width=0.45\linewidth}}
   {\epsfig{figure=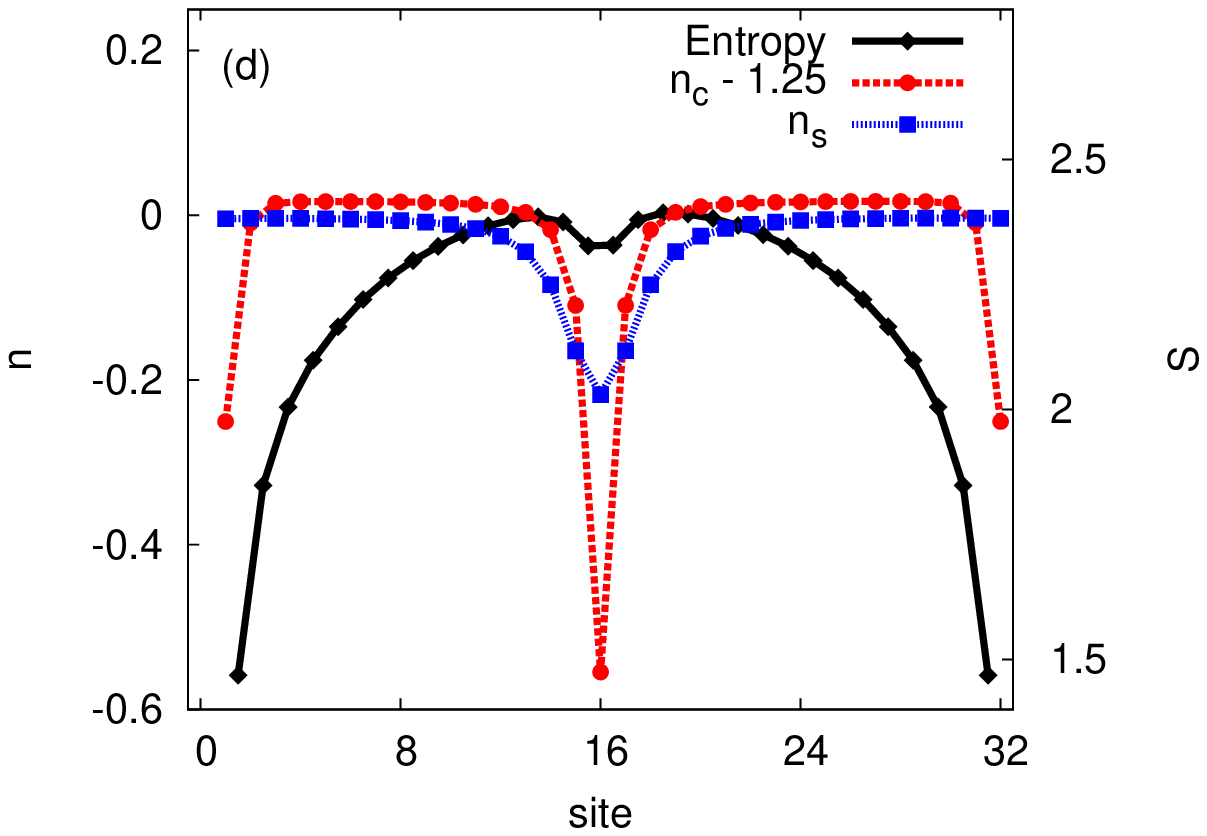,width=0.45\linewidth}} \\
   {\epsfig{figure=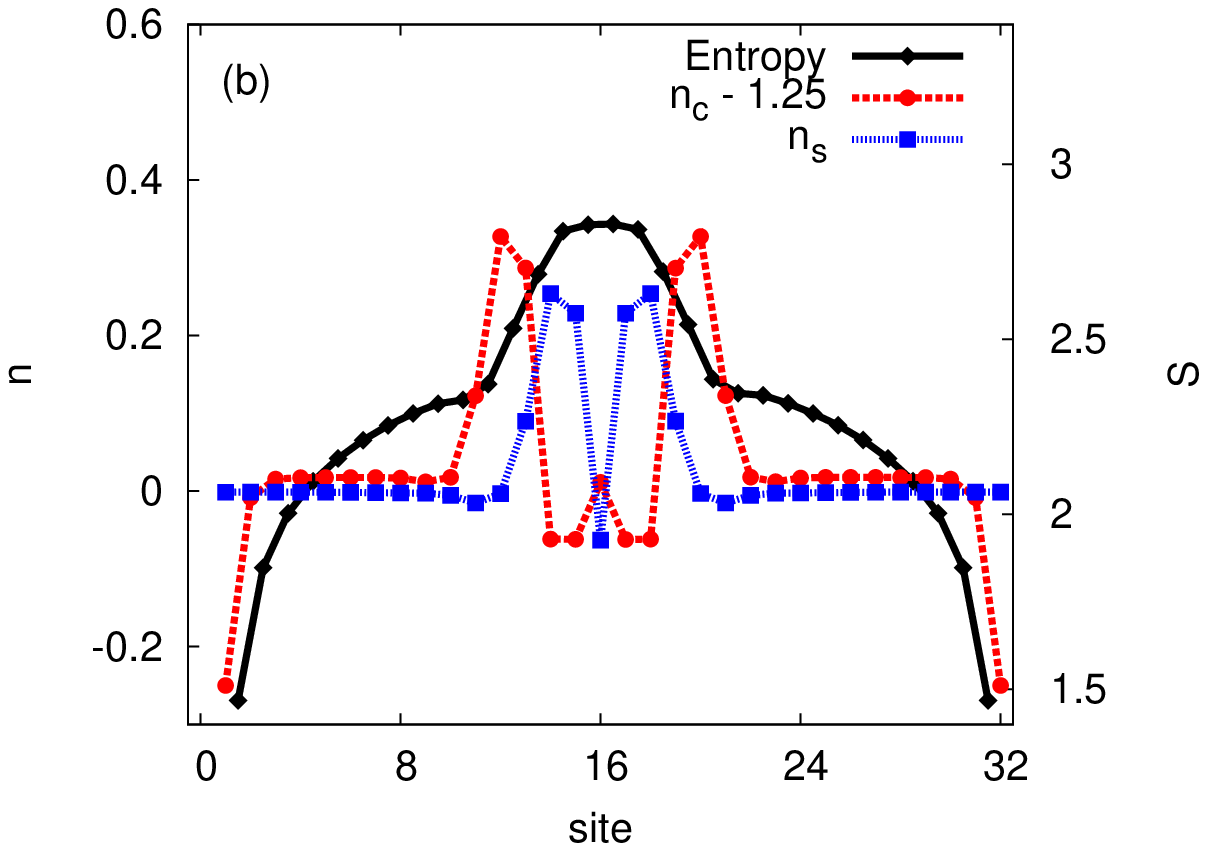,width=0.45\linewidth}}
   {\epsfig{figure=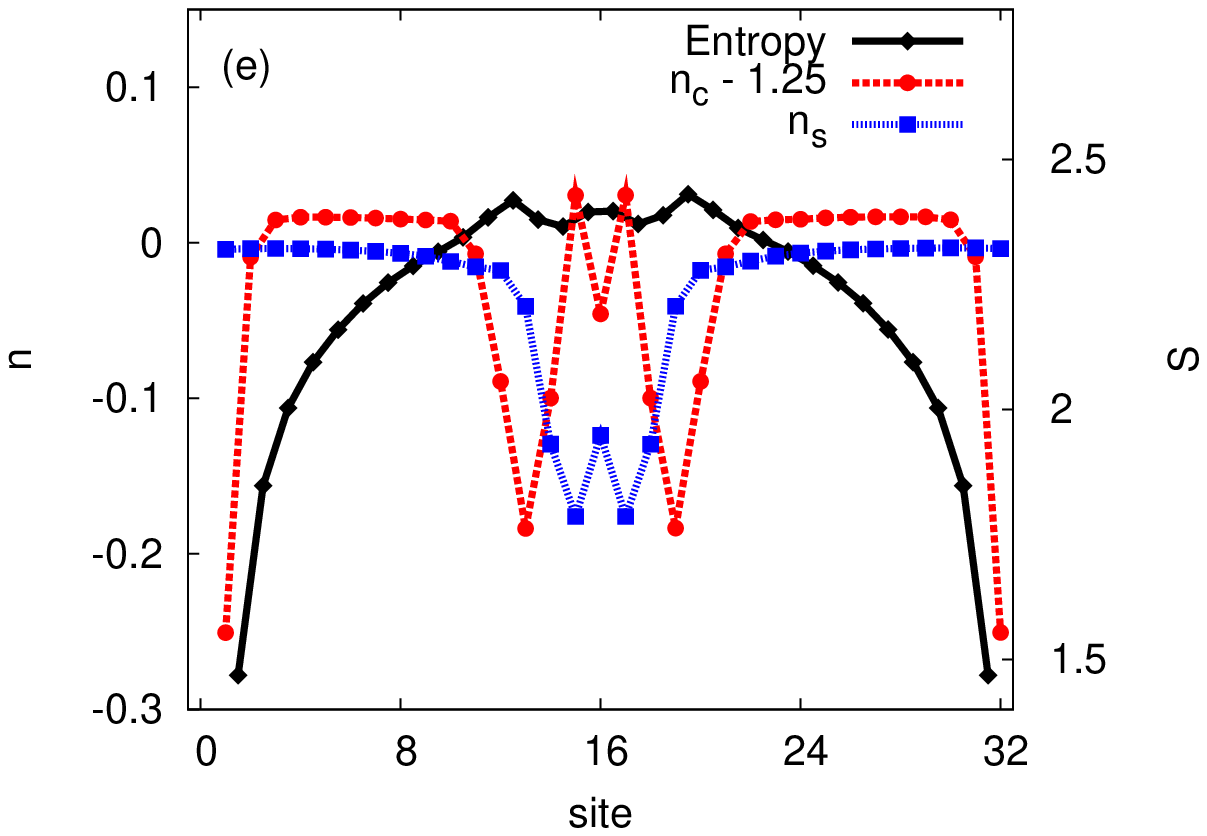,width=0.45\linewidth}} \\
   {\epsfig{figure=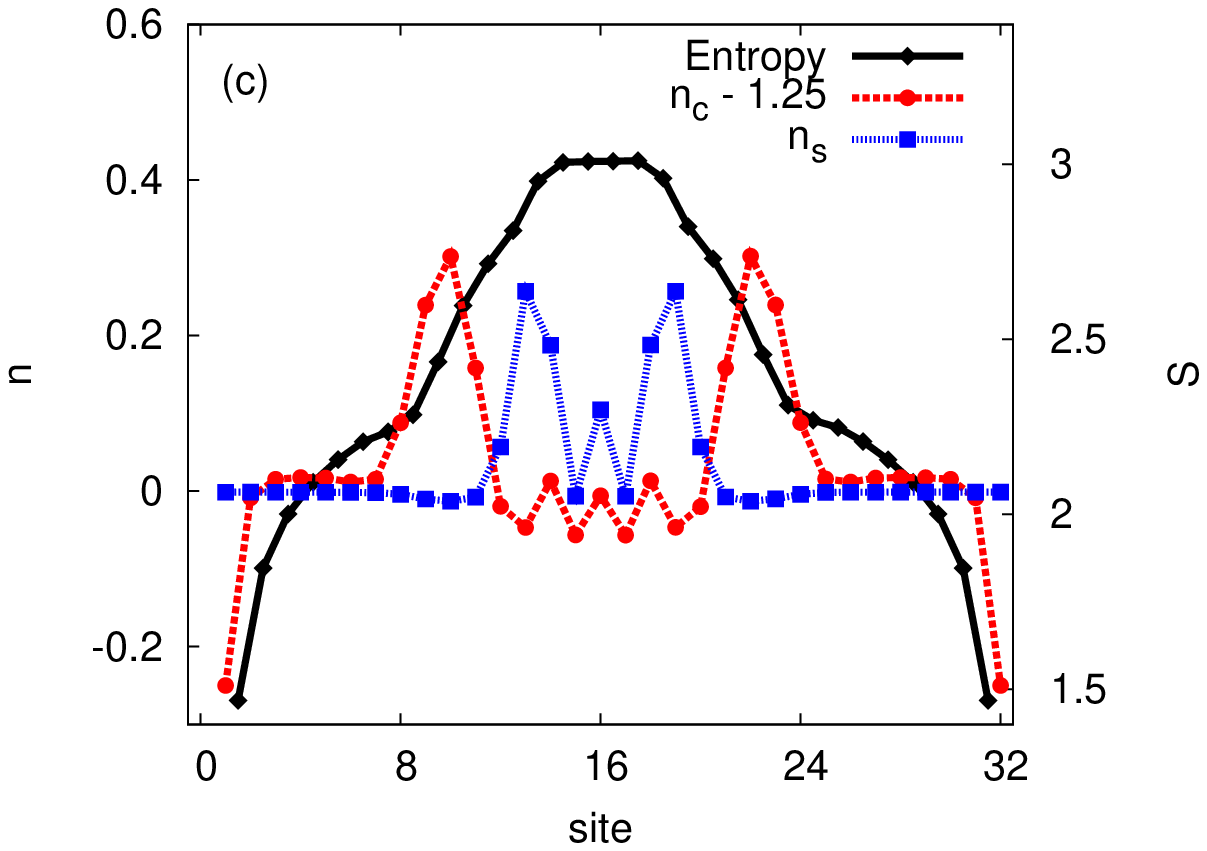,width=0.45\linewidth}}
   {\epsfig{figure=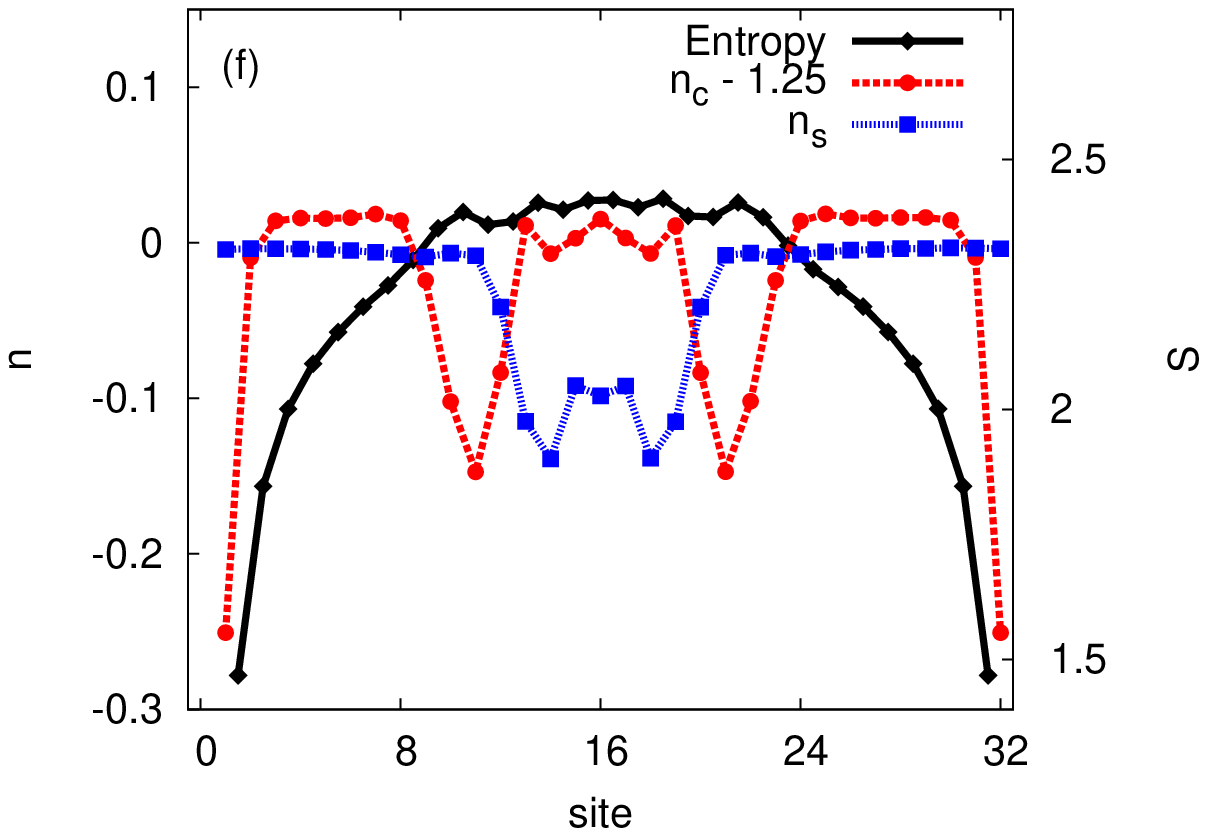,width=0.45\linewidth}}
\end{center}
\caption{(Color online) Snapshots of the time-evolution of the charge and spin density
 distribution of a single particle excitation ((a) - (c) creation of a particle, (d)-(f) annihilation of a particle)
 created at time $t=0\hbar/J$; (a), (d) at time $t=0\hbar/J$, (b), (e) at time $t=1.5 \hbar/J$ and (c), (f) at time $t=2.5
 \hbar/J$. The system parameters are $n_{1,2} = 0.625$, $U_1/J = U_2/J = 3.$, $U_{12} / J = 2.1$.
 The charge density is shifted by $1.25$ for better
 visibility.}
\label{fig:single_equal_b}
\end{figure}

\label{sec:spe}
The results for a density perturbation connect directly to a setup where the time-evolution of densities is followed upon the creation of single-particle excitations in bosonic gases instead of a density perturbation. In a one-dimensional system these single particle
excitations decay into a charge excitation and a spin excitation. In contrast
in a higher-dimensional system single particle excitations have a finite
life-time. Since in cold atomic gases one can realize systems of different
dimensionality, this second setup, if realized, would give the possibility to directly 
confront the
decay in a one-dimensional system with the finite life-time in a three
dimensional system. 
Fig.~\ref{fig:single_equal_b} shows the time evolution of the density and bipartite entanglement entropy profiles for a creation and an annihilation of a particle at time
$t = 0$. The initial excitation splits up into a right and a left moving part and one can clearly see the two different velocities for the
charge and spin density. This eventually leads to the separation into a spin
and a charge excitation. The entropy profiles exhibit a significant difference between the creation and
the annihilation of a particle. After removing one particle the entropy stays almost constant (up to some small waves) with respect to time, the entropy strongly
increases between the two counterprogating peaks after adding an addional particle. The reason for this behaviour is unknown; observations and conjectures will be discussed in
more detail in Sec.~\ref{sec:entropy}.

\section{Experimental constraints: Effects of unequal intraspecies interaction strengths and confining trap potentials}

\begin{figure} 
\begin{center}
   {\epsfig{figure=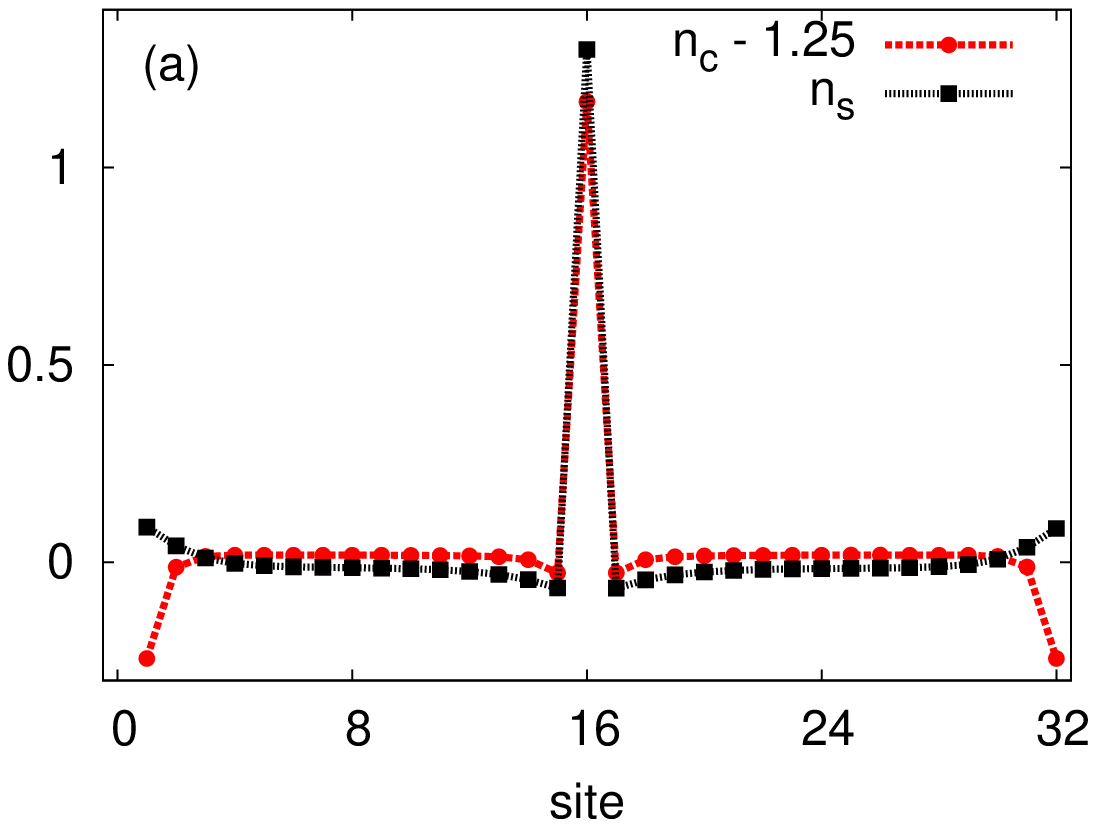,width=0.3\linewidth}}
   {\epsfig{figure=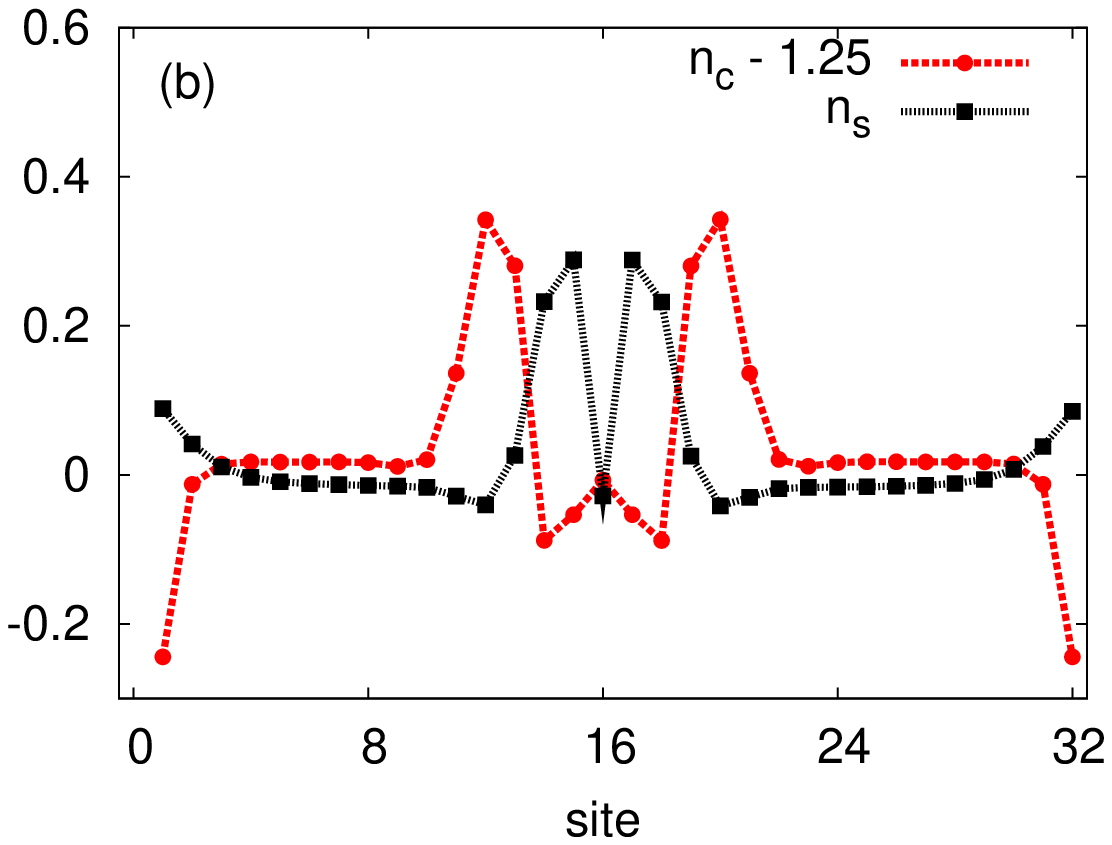,width=0.3\linewidth}}
   {\epsfig{figure=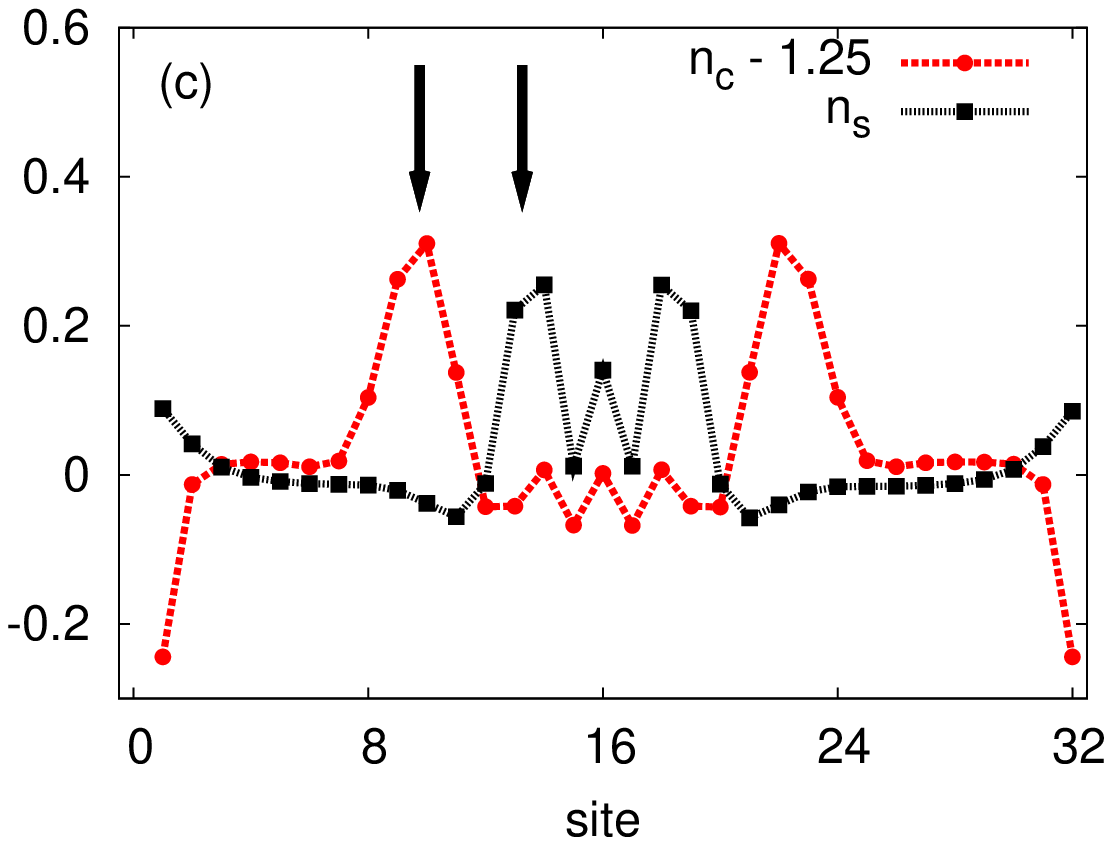,width=0.3\linewidth}}
\end{center}
\caption{(Color online) Snapshots of the time-evolution of the charge and spin density
 distribution of a single particle excitation created at time $t=0\hbar/J$; (a)
 at time $t=0\hbar/J$, (b) at time $t=1.5 \hbar/J$ and (c) at time $t=2.5
 \hbar/J$. The system parameters are $n_{1,2} = 0.625$, $U_1/J = 2.983$, $U_2/J = 2.712$, $U_{12} / J = 2.377$.
 The charge density is shifted by $1.25$ for better
 visibility. The arrows in (c) mark the clear separation of the charge and the spin
 density waves [Reproduced from Ref.~\cite{KleineSchollwoeck2007}], Copyright American Physical Society}
\label{fig:single_unequal}
\end{figure}

An experimental realisation of a two-component Bose-Hubbard model is given by the use of two different hyperfine states of $^{87}$Rb, for instance
$\ket{F=2, ~m_F = -1}$ and $\ket{F=1, ~m_F = 1}$. The $s$-wave scattering lengths for these states are approximatly $a_2 = 91.28 a_B$,
$a_1 = 100.4 a_B$, where $a_B$ is the Bohr radius \cite{WideraBloch2006}. 
Hence the intra-species interactions are slightly distinct for the
two bosonic species. By comparison, the interspecies scattering length $a_{12}$ is of the same order of magnitude and can be tuned by a Feshbach
resonance \cite{ErhardSengstock2004,WideraBloch2006}. Treating unequal intraspecies interactions in the bosonization approach results in a coupling of the spin and the charge part of the Hamiltonian. To decide if for the experimentally relevant parameters the coupling is already important we have calculated the time-evolution of a single particle excitation of a system with realistic parameters. The parameters have been determined using a lattice depth of $V_0 = 4.3 E_R$ and a interspecies scattering length $a_{12} = 80 a_B$, where $E_R$ is the recoil energy of the optical lattice. This yields the following Bose-Hubbard parameter,
$U_1/J = 2.983$, $U_2/J = 2.712$, $U_{12} / J =
2.377$. Fig.~\ref{fig:single_unequal} shows the density profiles for different
times. Even though a small coupling between the spin and the charge degree of
freedom might be present, the single particle excitation splits up into a
charge and a spin excitation. Therefore the separation in spin and charge
survives a small experimental mismatch in the interspecies interaction strength. 

A further complication of the ultracold atomic gases setup is the presence of
an parabolic trapping potential. In the Bose-Hubbard model this trapping
potential can be described by adding $\epsilon_{j,\nu} = V_0 \left(j - j_0\right)^2$. 

\begin{figure} 
\begin{center}
  {\epsfig{figure=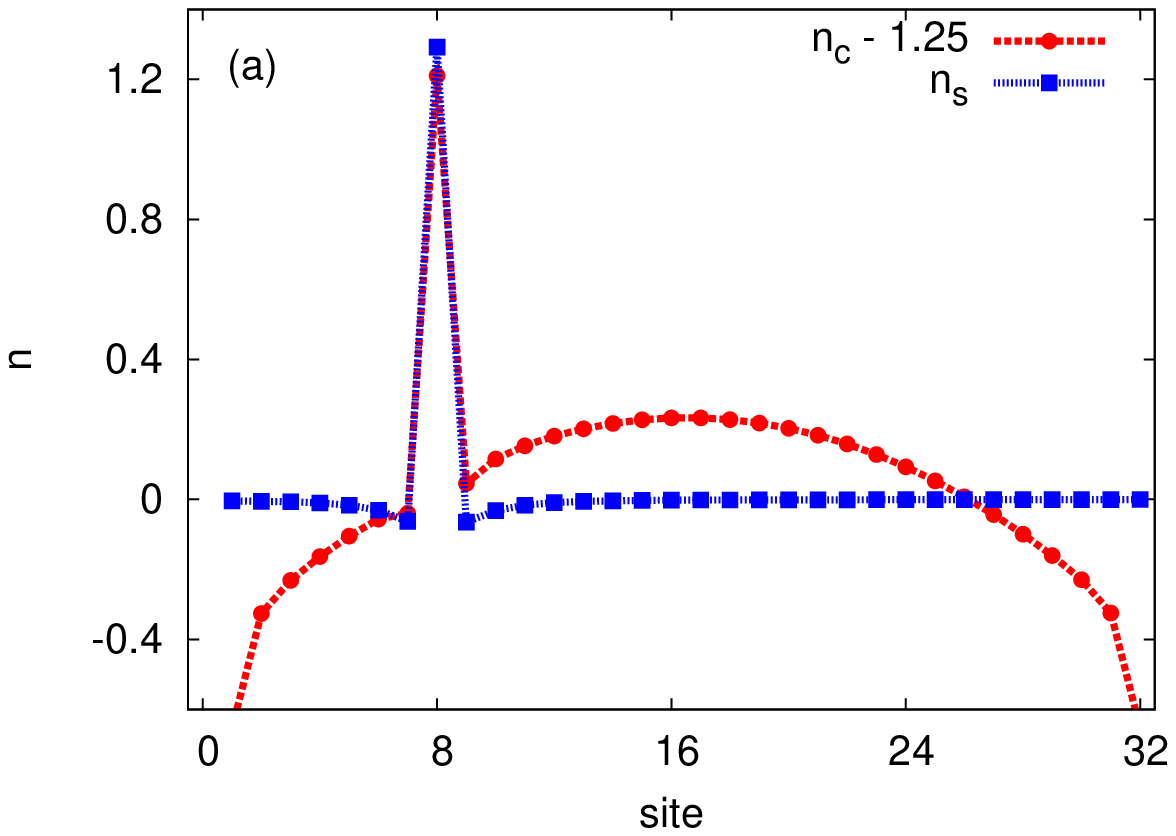,width=0.3\linewidth}}
  {\epsfig{figure=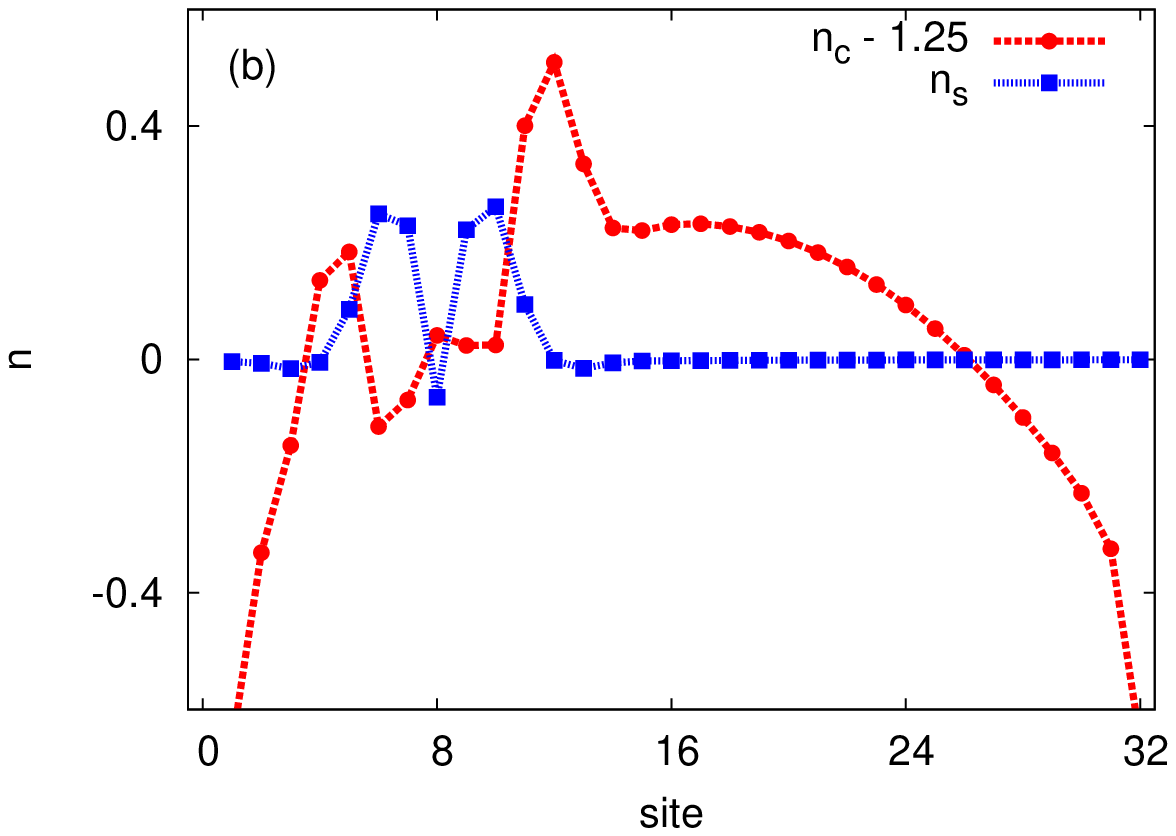,width=0.3\linewidth}}
  {\epsfig{figure=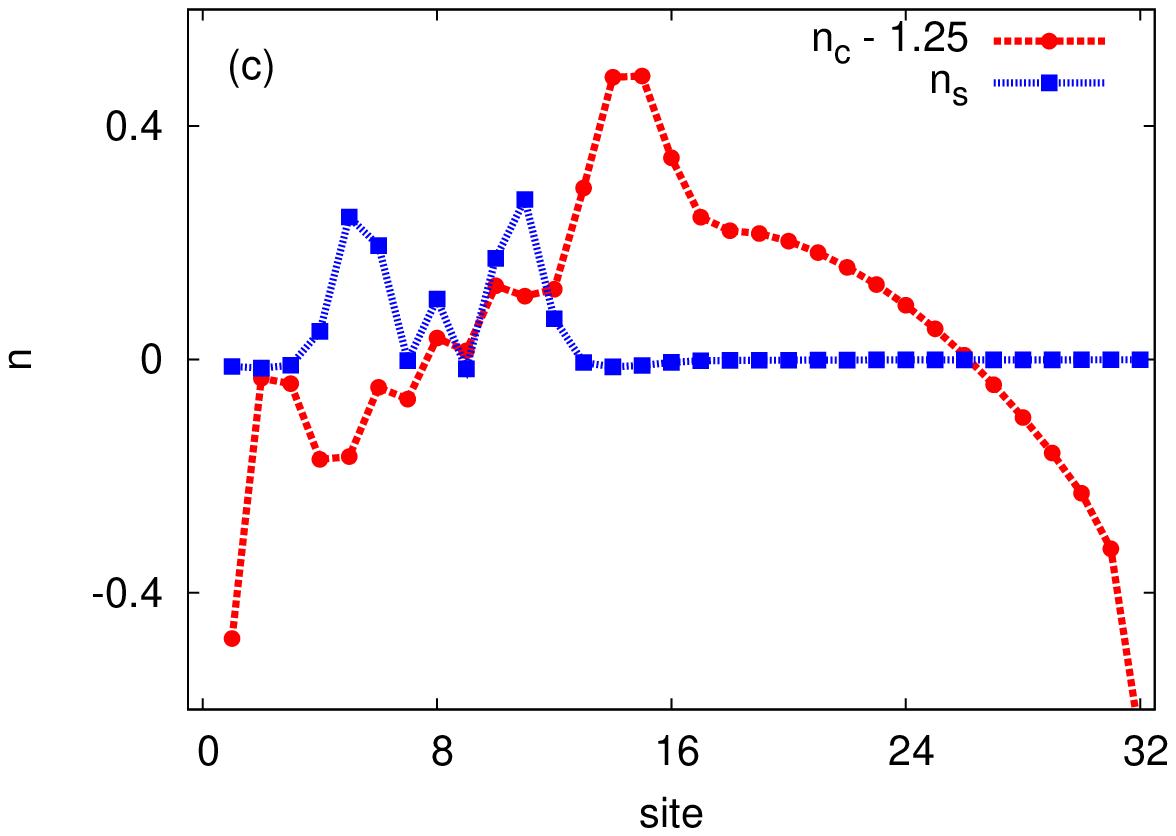,width=0.3\linewidth}}
\end{center}
\caption{(Color online) Snapshots of the time-evolution of the charge and spin density
 distribution of a single particle excitation created at time $t=0\hbar/J$ in a trap; (a)
 at time $t=0\hbar/J$, (b) at time $t=1.5 \hbar/J$ and (c) at time $t=2.5
 \hbar/J$. The system parameters are $n_{1,2} = 0.625$, $u_1 = 3$, $u_{12} = 2.1$ and
 $\epsilon_{j,\nu} = 6\cdot 10^{-3} \left( j - j_0 \right)^2 $.
 The charge density is shifted by $1.25$ for better
 visibility.}
\label{fig:single_trap}
\end{figure}

We calculated the time-evolution of a single particle excitation with a trapping potential, see Fig.~\ref{fig:single_trap}. In contrast to the case without a trap
the ground state density is not constant anymore. Therefore the velocity of the
charge and the spin excitation now depend on the spatial position of the
excitation. Still, two counter-propagating waves can be observed and the spin-charge separation is not qualitatively changed. The effect of a trapping potential has already been discussed in the context of spin-charge separation for two component fermionic systems \cite{RecatiZoller2003,KeckeHaeusler2004,KollathZwerger2005}, where similar robustness of the results was found.

\section{Entropy of entanglement}
\label{sec:entropy}
The superposition of states that is characteristic of quantum mechanics implies for many-body systems the phenomenon
of entanglement that is the key deviation of the quantum from the classical world and a key
resource of quantum computing. In this Section, we focus on one measure of entanglement in bipartite systems, the entropy
of entanglement, which is defined in the case of pure quantum states for an arbitrary bipartition of the system into a
left part A and a right part B by cutting at bond $i$ by forming the reduced density operators of A and B,

\begin{equation}
\label{eq:reduced_density_matrix}
\hat \rho_A = \Tr_B \left( \ket{\Psi}\bra{\Psi} \right) \quad\quad \hat \rho_B = \Tr_A \left( \ket{\Psi}\bra{\Psi} \right).
\end{equation}

The entropy of entanglement is then given by the von Neumann entropy of either reduced density operator $\hat \rho_A$ or $\hat \rho_B$:

\begin{equation}
\label{eq:van-neumann_entropy}
S =  - \Tr_A \left( \hat\rho_A\log \hat\rho_A \right) = - \Tr_B \left( \hat\rho_B\log \hat\rho_B \right) .
\end{equation}
The identity of both definitions follows from the well-known observation that the eigenspectra of reduced density operators of a bipartition are identical.

Entropy of entanglement has been studied in numerous contexts. In the present work, the focus is on the time-evolution of the entropy
of entanglement in an out-of-equilibrium setting. On the one hand, this question is of fundamental interest for the understanding of
coherent out-of-equilibrium quantum dynamics. On the other hand, it turns out that the time-evolution of the entropy of entanglement
is closely related to the performance of time-dependent DMRG and related methods: the number of states (matrix dimensions) needed in
such simulations are in a roughly exponential relationship with the entropy of entanglement. This means that e.g.\ a linear growth of
entanglement entropy in time is reflected in an exponential growth in matrix dimensions, yielding the key limitation\cite{Gobert2004}
for the time-scales accessible for such algorithms. 

Building on the Lieb-Robinson theorem\cite{LiebRobinson1972} it has been shown by Osborne\cite{Osborne2006} that for arbitrary
short-ranged Hamiltonians in one dimension, entanglement growth is bounded linearly in time, $S(t) \leq S(0) + ct$, reflecting
a finite speed of propagation in such Hamiltonians, implying a potentially exponentially growth of matrix dimensions in time. In fact,
such linear growth of entanglement has been observed\cite{CalabreseCardy2004,CalabreseCardy2005,Chiara2006} and been traced back to the
fact that ouf of equilibrium quantum states show excitations propagating through the system, leading to a linearly expanding "light cone"
where information is exchanged between bipartition parts A and B.

In this paper, we are considering three quite different types of time-evolution: (i) time-evolution after an insertion of a single particle,
(ii) time-evolution after the extraction of a single particle, (iii) time-evolution after removing an external potential that created a density
perturbation. All processes should generate excitations moving at the same maximal finite speed of propagation; one would therefore naively expect
that all cases lead to a qualitatively similar linear growth of entanglement entropy in time, although we cannot expect identical results: for
example, there is no particle-hole symmetry relating (i) and (ii).

\begin{figure} 
\begin{center}
{\epsfig{figure=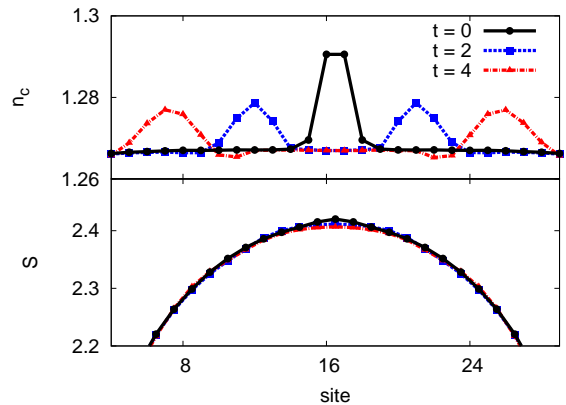,width=0.5\linewidth}}
\end{center}
\caption{(Color online) Entropy for a density perturbation $n_{1,2} = 0.625$, $U_1/J = U_2/J = 3.$, $U_{12} / J = 2.1$.}
\label{fig:entropy_small}
\end{figure}

Fig.~\ref{fig:single_equal_b} shows the time-evolution of the entropy profiles after a single particle excitation by either insertion (left) or removal (right),
i.e.\ cases (i) and (ii) where the entanglement entropy has been measured for all possible bipartitions. The drops at the ends are a natural consequence of
the small dimension of either $\hat{\rho}_A$ or  $\hat{\rho}_B$ there. The difference between (i) and (ii) is striking. In the case (i) entanglement entropy
increases roughly linearly in time, as expected. However, in case (ii) entanglement entropy stays almost constant. This implies that simulating case (i)
is much harder numerically than case (ii). For a comparison, we show case (iii), a small density perturbation evolving in time: the entanglement entropy
is almost entirely unchanged under the time-evolution, see Fig.~\ref{fig:entropy_small}. \\

\begin{figure} 
\begin{center}
   {\epsfig{figure=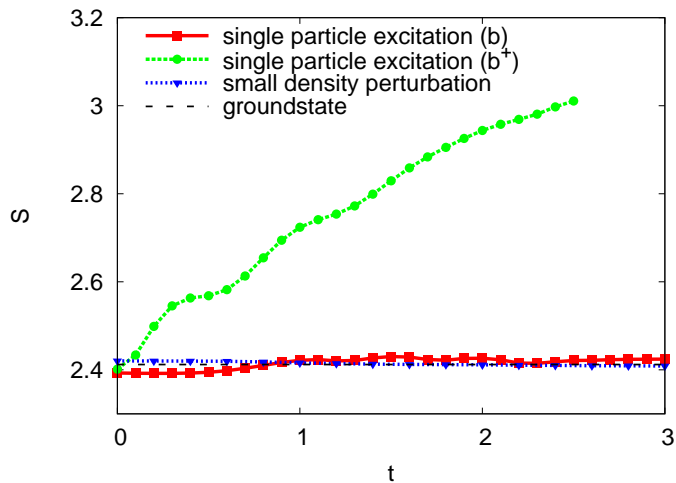,width=0.6\linewidth}}
\end{center}
\caption{(Color online) Maximum of the bond entropy versus time for single particle excitations (both $b^\dagger\ket{0}$ and $b\ket{0}$)
  and for a small density perturbation. The small wave occuring for the $b^\dagger\ket{0}$ curve are an effect of the lattice.
  Note the enormous difference between the two single particle excitations.
  The system parameters are $n_{1,2} = 0.625$, $u = 3$, $u_{12} = 2.1$.}
\label{fig:single}
\end{figure}

To summarise these observations, we show the maximal entanglement entropy on a chain versus time for cases (i)-(iii) in Fig.~\ref{fig:single},
with the ground state entanglement given as a reference. As already seen in Fig.~\ref{fig:single_equal_b}, a enormous difference between the creation
and the annihilation can be observed. 

We can exclude that these observations are due to limitations of the numerical method that obviously neglects some of the entanglement entropy due
to its inherent truncations. However, these results are converged in the used matrix dimensions; we also have calculated the fidelities between
the initial states and the result of time-evolutions up to time $t$ and then back to $0$, finding fidelities that are essentially 1. The
entanglement entropies shown can therefore be considered exact. 

In the case of a weak density perturbation, we attribute the observed behaviour to the fact that the out-of-equilibrium wave function is essentially
a weakly distorted ground state wave function that does not affect the entanglement structure of the ground state substantially. It is much less
obvious to explain the difference between (i) and (ii) despite the absence of a particle-hole symmetry. If we consider Fig.~\ref{fig:single_equal_b},
in the case of particle creation the increase of entanglement entropy is limited to the regions into which spin and charge excitations have already
propagated that link ever larger regions of space. This is as expected. In the case of particle annihilation we might suppose that due to the relatively
low density entanglement is removed because the chain is to some extent cut by
the removal of a particle. In particular, if one thinks of the initial
state as a superposition of many different Fock states, the application of
the annihilation operator eliminates all states in which no particle occupies the site. Thereby the entanglement in the system
is reduced. Indeed, a small dip is observed. The
perturbation in the entanglement entropy propagates again linearly with the charge and spin excitations, but quantitatively it stays essentially
unchanged. A possible explanation is suggested by perturbation theory.
For short times the time-evolution operator can be approximated by an expression of the structure $1+\textrm{i} t J \sum (b^\dagger_{i}b_{i\pm 1} + h.c.) + \textrm{i} t (U/2) n_{i}(n_{i}-1)$, where we have suppressed the multi-component nature of the problem for illustrative purposes. While the time-evolution operator applied far from the perturbation conserves the state and thereby
the entanglement entropy, we see that close to the perturbation changes
occur. At the densities considered, the site where the annihilation operator is applied has most likely occupation number 0 or 1 for the respective species in the contributing Fock states. After application of the operator, the Fock states either vanish or have occupation number 0 on the site, such that only one of the terms in the hopping operator couples
to the state at the site of the perturbation. Rerunning the argument for the creation operator, no contribution vanishes (if we assume intermediate interaction strengths) and all hopping terms couple, which might indicate that changes in entanglement are much more pronounced in the latter case, as observed. This is obviously not rigorous at all, and at the moment
we can only conclude that the simple picture of excitations transporting entanglement leading to linear entanglement growth needs substantial refinement.

\ack

We would like to thank J.S.~Caux, S. F\"olling, M. K\"ohl, B. Paredes and A. Kolezhuk for fruitful
discussions. AK and US acknowledge support by the DFG (FOR 801) and CK and TG
by the Swiss National Science Foundation under MaNEP and
Division II and the CNRS.

\appendix
\section{Single-species Bose-Hubbard model}
\begin{figure}
\begin{center}
{\epsfig{figure=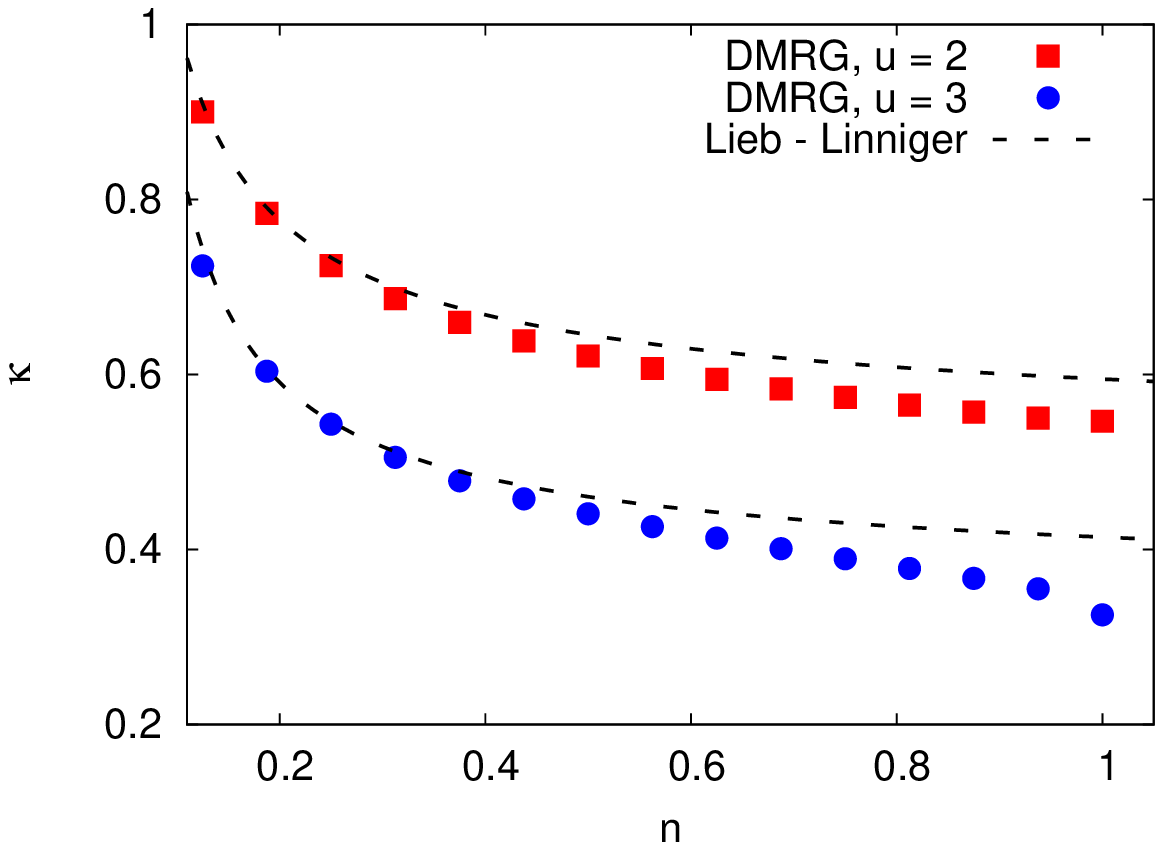,width=0.5\linewidth}}
\end{center}
\caption{(Color online) Compressibility $\kappa = K / v\pi$ of a one-component Bose - Hubbard model for $u = 2$ and $u = 3$.}
\label{fig:Lutt_BH}
\end{figure}

The single-species Bose-Hubbard model has been discussed extensively in
previous literature; for an overview of the literature, see \cite{BlochDalibardZwerger2007}. In this appendix we would
like to demonstrate some technical details useful to understand our results
for the two-species Bose-Hubbard
model in more detail at the example of the one-species model. 
The Hamiltonian of the one-species Bose-Hubbard model is given by
\begin{equation}
\label{eq:H_BHM}
\begin{array}{rl}
H =& -J \sum_{j} \left( b^\dagger_{j+1}b_{j,} +
h.c.\right) + \sum_{j} \frac{U \hat{n}_{j}(\hat{n}_{j}-1)}{2}  \\
&+ \sum_{j}
\varepsilon_{j} \hat{n}_{j}, 
\end{array}
\end{equation} 
where we used the same notation as for the two-component model. 
Let us first comment on the deviation we see comparing the values of the
velocities and the compressibility with the analytical formulae for the
continuum limit. 
For the single component Bose-Hubbard model the expressions are given by
\begin{equation}
\label{eq:Kv_LL}
\begin{array}{rl}\kappa_0 = & K_0/(\pi v_0)\\
K_0 = & \frac{\pi}{\sqrt{\gamma}} \left( 1 - \sqrt{\gamma} / (2\pi) \right)^{-\frac12} \\
v_0 = & 2n\sqrt{\gamma} \left( 1 - \sqrt{\gamma} / (2\pi) \right)^{\frac12}
\end{array}
\end{equation} 
with interaction parameter $\gamma=U/2n$. In Fig.~\ref{fig:Lutt_BH} we compare the numerical results for the
compressibility $\kappa_0 = K_0/(\pi v_0)$ with the analytical formula for the
continuum limit. The Lieb-Liniger results yield
a good approximation for the compressibility for small densities $n$. This corresponds to rather high values of $\gamma$ of the order of $10$. In
contrast, for larger densities above $0.6$ the results deviate considerably even for small values of $\gamma \approx 1$. Kollath et al. showed in \cite{KollathZwerger2004}
that the velocity of small density perturbation of the Bose-Hubbard model is
given by the Lieb-Liniger solution for $n < 1$ and $\gamma$ up to
$4$. Therefore we see the same effect as in the case for the two-component
model, that the deviations for the compressibility at $n\approx 1$ become
larger than the deviations for the velocities.\\

\begin{figure} 
\begin{center}
{\epsfig{figure=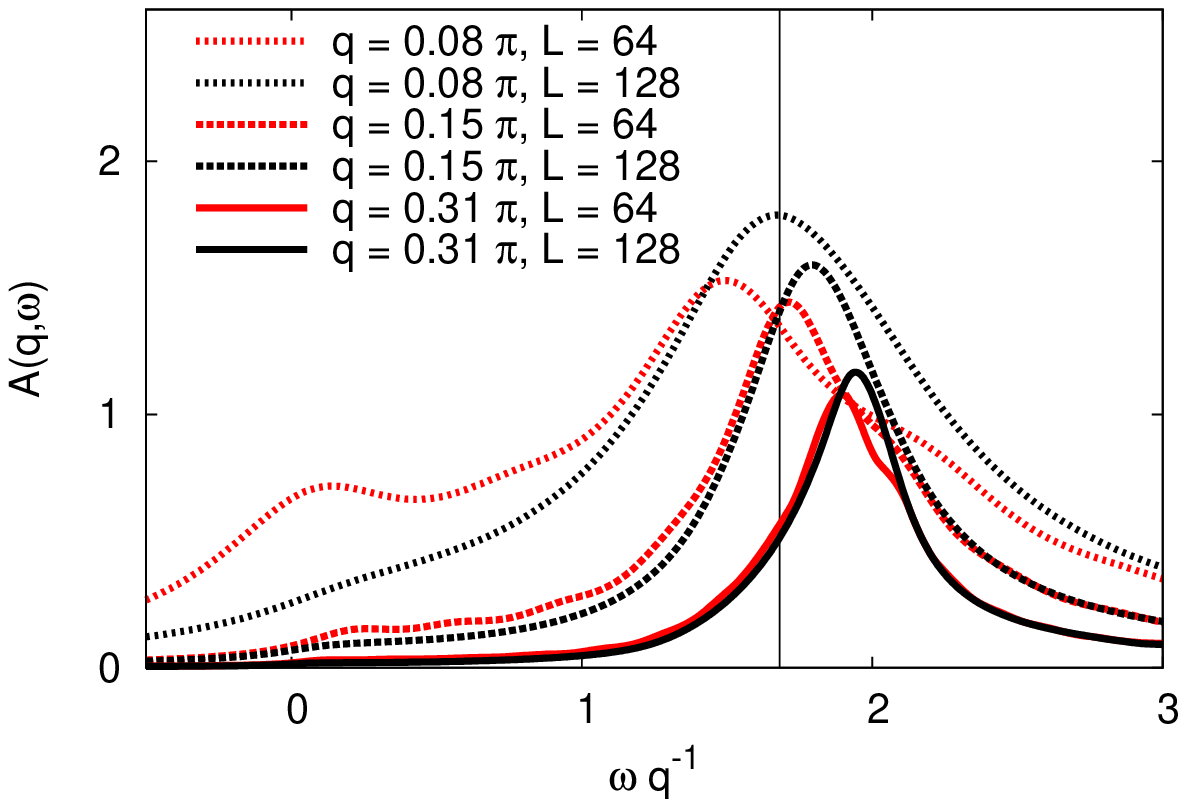,width=0.5\linewidth}}
\end{center}
\caption{(Color online) Finite size scaling of the one-particle spectral function $A(q,\omega)$ at momenta $q = 5/65 \pi/a$ and $q = 10/129 \pi/a$
 respectively for a one-component Bose-Hubbard model.
 The vertical lines mark the   position of $u_0$. The following parameters
 were used $n = 0.625$, $u = 3$ on a system with
 $L=64$ and $L=128$ sites and a broadening $\eta=0.1$. }
\label{fig:spectral_BH}
\end{figure}

To investigate finite-size effects on the single-particle spectral function we
show here the results for the single-component Bose-Hubbard model.
Two different system lengths were considered in order to explore the boundary effects
of a finite-size system. In fig.~\ref{fig:spectral_BH} we show the spectral function $A(q,\omega)$ as a function of $\omega / q$. Assuming a linear dispersion relation,
one would expect a peak of the spectral function at the velocity $v_0$, see \cite{MedenSchoenhammer1992}. 
The numerical results indeed show this behaviour. Furthermore, a number of boundary effects are visible: At
$\omega \approx 0$ there exists an additional peak and the main peak is
shifted to lower frequencies. Above the main peak there is also an additional shoulder. These are the same effects as we find for the
two-component model. Let us note that all boundary
effects become less pronounced for larger values of $q$.

%\section*{References}
%\bibliographystyle{unsrt}
%\bibliography{references}

\end{document}